\documentclass[12pt,a4paper]{article}
\usepackage{jheppub}

\usepackage{graphicx, epsfig}
\usepackage{color}
\usepackage{mathrsfs}
\usepackage{amsmath}

\newcommand{\order}{\mathcal{O}}
\newcommand{\Vol}{\hat{\mathcal{V}}}
\newcommand{\be}{\begin{equation}}
\newcommand{\ee}{\end{equation}}
\newcommand{\bea}{\begin{eqnarray}}
\newcommand{\eea}{\end{eqnarray}}
\newcommand{\barr}{\begin{array}}
\newcommand{\earr}{\end{array}}

\def\beq{\begin{equation}}
\def\eeq{\end{equation}}
\def\be{\begin{equation}}
\def\ee{\end{equation}}
\def\bea{\begin{eqnarray}}
\def\eea{\end{eqnarray}}

\title{A sufficient condition for de Sitter vacua in type IIB string theory}
\author[a]{Markus~Rummel} \author[b]{and Alexander~Westphal}

\affiliation[a]{II. Institut f\"ur Theoretische Physik der Universit\"at Hamburg, D-22761 Hamburg, Germany}
\affiliation[b]{Deutsches Elektronen-Synchrotron DESY, Theory Group, D-22603 Hamburg, Germany}
\emailAdd{markus.rummel@desy.de}\emailAdd{alexander.westphal@desy.de}
\abstract{We derive a sufficient condition for realizing meta-stable de Sitter vacua with small positive cosmological constant within type IIB string theory flux compactifications with spontaneously broken supersymmetry. There are a number of `lamp post' constructions of de Sitter vacua in type IIB string theory and supergravity. We show that one of them -- the method of `K\"ahler uplifting' by F-terms from an interplay between non-perturbative effects and the leading $\alpha'$-correction -- allows for a more general parametric understanding of the existence of de Sitter vacua. The result is a condition on the values of the flux induced superpotential and the topological data of the Calabi-Yau compactification, which guarantees the existence of a meta-stable de Sitter vacuum if met. Our analysis explicitly includes the stabilization of all moduli, i.e. the K\"ahler, dilaton and complex structure moduli, by the interplay of the leading perturbative and non-perturbative effects at parametrically large volume.
}

\begin{document}
\noindent July 11, 2011\hfill
DESY 11-121
\\[0.7cm]
\maketitle
\section{Introduction \& Motivation}

\setlength{\parskip}{0.2cm} 

String theory is a candidate for a fundamental theory of nature, providing at the same time a UV-finite quantum theory of gravity and unification of all forces and fermionic matter. Mathematical consistency requires string theory to live in a ten dimensional space-time, and a description of our large four-dimensional physics thus necessitates compactification of the additional six dimensions of space.

The need for compactification confronts us with two formidable consequences: Firstly, even given the known internal consistency constraints of string theory, there are unimaginably large numbers of 6d manifolds available for compactification. Secondly, many compact manifolds allow for continuous deformations of their size and shape while preserving their defining properties (such as topology, vanishing curvature, etc) -- these are the moduli, massless scalar fields in 4d. This moduli problem is exacerbated if we wish to arrange for low-energy supersymmetry in string theory, as compactifications particularly suitable for this job -- Calabi-Yau manifolds -- tend to come with hundreds of complex structure and K\"ahler moduli.

Therefore, a very basic requirement for string theory to make contact with low-energy physics is moduli stabilization -- the process of rendering the moduli fields very massive. Moreover, as supersymmetry is very obviously broken -- and so far has not been detected -- ideally, moduli stabilization should tolerate or even generate supersymmetry breaking. And finally, the process should produce a so-called meta-stable de Sitter (dS) vacuum with tiny positive cosmological constant, so as to accommodate the observational evidence for the accelerated expansion of our universe by dark energy~\cite{Perlmutter:1998np,Riess:1998cb,Komatsu:2010fb}.

The task of moduli stabilization and supersymmetry breaking has recently met with considerable progress, which is connected to the discovery of an enormous number~\cite{Bousso:2000xa,Giddings:2001yu,Kachru:2003aw,Susskind:2003kw,Douglas:2003um} of stable and meta-stable 4d vacua in string theory. The advent of this �landscape�~\cite{Susskind:2003kw} of isolated, moduli stabilizing minima marks considerable progress in the formidable task of constructing realistic 4d string vacua.

There are several methods of moduli stabilization. The first one uses supersymmetric compactifications of string theory on a Calabi-Yau manifold, and the strong gauge dynamics of gaugino condensation in the `racetrack' mechanism to stabilize the dilaton and several of the bulk volume and complex structure moduli~\cite{Derendinger:1985kk,Dine:1985rz,deCarlos:1992da}. Recently, this method has been applied to supersymmetric compactifications of M-theory on $G_2$-manifolds, where the structure of the manifolds allows for the racetrack superpotential to generically depend on all the moduli of the compactification~\cite{Acharya:2006ia}.

The second, more recent, method relies on the use of quantized closed string background fluxes in a given string compactification. These flux compactifications can stabilize the dilaton and the complex structure moduli of type IIB string theory compactified on a Calabi-Yau orientifold supersymmetrically~\cite{Giddings:2001yu}. The remaining volume moduli are then fixed supersymmetrically by non-perturbative effects, e.g. gaugino condensation on stacks of D7-branes~\cite{Kachru:2003aw}. The full effective action of such fluxed type IIB compactifications on Calabi-Yau orientifolds was derived in~\cite{Grimm:2004uq}. In type IIA string theory on a Calabi-Yau manifold all geometric moduli can be stabilized supersymmetrically by perturbative means using the larger set of fluxes available~\cite{DeWolfe:2005uu}.

If the moduli are stabilized supersymmetrically, parametrically small and controlled supersymmetry breaking can happen, e.g, by means of inserting an anti-D3-brane into a warped throat of the Calabi-Yau~\cite{Kachru:2003aw}, by D-terms originating in magnetic flux on a D7-brane~\cite{Burgess:2003ic}, or dynamically generated F-terms of a matter sector~\cite{Lebedev:2006qq}. This process is known as `uplifting' and allows for dS vacua with extremely small vacuum energy by means of fine-tuning the ${\cal O}(100)$ independent background fluxes available in a typical Calabi-Yau compactification~\cite{Bousso:2000xa,Kachru:2003aw}. The reliability of this last step of uplifting supersymmetric AdS vacua without unstabilized moduli into a dS vacuum is still under discussion. Some of the points in question e.g. concern the fact that the existence of D7-brane D-terms as well as F-terms from hidden matter sectors are very model dependent, rendering statistical sweeps over large sets of compactifications difficult. Supersymmetry breaking and uplifting by a warped-down anti-D3-brane also remains under ongoing discussion on whether its presence can be completely described in a probe approximation or causes dangerous non-normalizable perturbations to the compact geometry~\cite{DeWolfe:2008zy,Bena:2009xk,Dymarsky:2011pm,Blaback:2011nz,Bena:2011wh,Burgess:2011rv}. Very recently, the use of internal $F_2$ gauge flux on a CY threefold in heterotic string theory has been used to stabilize all geometric  moduli except the dilaton and one K\"ahler modulus in a supersymmetric Minkowski vacuum~\cite{Anderson:2010mh,Anderson:2011cz}.

Alternatively, in non-Calabi-Yau flux compactifications of type IIB or IIA string theory, all geometric moduli can be stabilized perturbatively in a non-supersymmetric way using a combination of background fluxes, D-branes, orientifold planes, and negative curvature. Examples here are flux compactifications of type IIB with 3-form fluxes on a product of Riemann surfaces~\cite{Saltman:2004jh} and almost Calabi-Yau 4-folds in F-theory~\cite{Dong:2010pm}, type II compactifications with generalized fluxes on manifolds of $SU(3)$ (see, e.g., the reviews~\cite{Grana:2005jc,Douglas:2006es}), as well as of type IIA with fluxes on a product of two 3d nil manifolds~\cite{Silverstein:2007ac}. The ingredients used typically lead to scalar potential dominated by three perturbative terms with alternating signs, which depend as varying power laws on the dilaton and the geometric moduli. Such a `3-term structure'  structure generically allows for tunable dS vacua~\cite{Saltman:2004jh,Silverstein:2007ac}. Supersymmetry is generically broken in these perturbative mechanisms of moduli stabilization at a high scale, which typically is the Kaluza-Klein (KK)-scale. The geometric and flux part of these type IIA compactifications were studied in more detail in~\cite{Caviezel:2008tf,Flauger:2008ad,Danielsson:2009ff,Douglas:2010rt,Wrase:2010ew,Danielsson:2010bc,Underwood:2010pm,Shiu:2011zt,Burgess:2011rv}. The conclusion there so far seems to be that in absence of the KK5-branes used in~\cite{Silverstein:2007ac} (which play a similar role as 'explicit' supersymmetry breaking objects as the anti-D3-brane in~\cite{Kachru:2003aw}) there are no stable dS vacua. A complete analysis including the effects of the KK5-branes in the language of~\cite{Caviezel:2008tf,Flauger:2008ad,Danielsson:2009ff,Douglas:2010rt,Wrase:2010ew,Danielsson:2010bc,Underwood:2010pm,Shiu:2011zt,Burgess:2011rv} still remains open.

Finally, in type IIB flux compactifications on Calabi-Yau manifolds there are constructions of a `hybrid' type, where fluxes fix the complex structure moduli and the dilaton supersymmetrically, but the volume moduli are stabilized non-super-symmetrically by an interplay of non-perturbative effects on D7-brane stacks and the leading perturbative correction at ${\cal O}(\alpha'^3)$ in type IIB~\cite{Becker:2002nn}, or by perturbative corrections to the K\"ahler potential alone.  Examples for the latter consist of the Large-Volume-Scenario (LVS)~\cite{Balasubramanian:2005zx}, stabilization by perturbative corrections to the K\"ahler potential of the volume moduli alone~\cite{Berg:2005yu,Berg:2007wt,Cicoli:2007xp,vonGersdorff:2005bf} which are uplifted by D7-brane D-terms~\cite{Parameswaran:2006jh}, and the method of `K\"ahler uplifting'~\cite{Westphal:2006tn,Balasubramanian:2004uy}.

For `K\"ahler uplifted' dS vacua, an interplay between the leading perturbative correction at ${\cal O}(\alpha'^3)$ and a non-perturbative effect in the superpotential serves to generate a dS vacuum with supersymmetry spontaneously broken by an F-term generated in the volume moduli sector.
For some recent reviews on flux compactifications and the associated questions of the landscape of string vacua and string cosmology ensuing from the meta-stable dS vacua, with a much more complete list of references, please see~\cite{Douglas:2006es,McAllister:2007bg,Baumann:2009ni}.

`K\"ahler uplifting' has the benefit of generating meta-stable dS vacua in terms of just background 3-form fluxes, D7-branes and the leading perturbative ${\cal O}(\alpha'^3)$-correction, data which are completely encoded in terms of the underlying F-theory compactification on a fluxed Calabi-Yau fourfold.   In addition, supersymmetry is spontaneously broken at a scale of order of the inverse Calabi-Yau volume, measured in string units this is typically $\sim M_{\rm GUT}$ here, and still below the KK-scale), by an F-term generated in the volume moduli sector. No extra anti-branes, D-terms or F-term generating matter fields are needed or involved. The existing analysis of these models consists of including manifestly the dilaton and one complex structure modulus~\cite{Westphal:2006tn}.

Therefore, in this paper we develop a method towards a rigorous analytical understanding of `K\"ahler uplifting' driven by the leading ${\cal O}(\alpha'^3)$ correction to the K\"ahler potential of the volume moduli. Our derivation will be carried out in the presence of an arbitrary number $h^{2,1}$ of complex structure moduli. A large value of 3-cycles $h^{2,1}={\cal O}(100)$  is a prerequisite to use the associated 3-form fluxes for the required fine-tuning of the cosmological constant.

Note the relationship between the supersymmetric KKLT-type AdS vacua~\cite{Kachru:2003aw} (prior to uplifting) with the flux superpotential tuned small, the SUSY-breaking LVS-type AdS vacua~\cite{Balasubramanian:2005zx} (again, prior to uplifting), and the SUSY-breaking `K\"ahler uplifted' AdS/dS vacua~\cite{Westphal:2006tn,Balasubramanian:2004uy} (inherently liftable to dS by the pure moduli sector itself) discussed here. These three classes of moduli stabilizing vacua are three branches of solutions in the same low-energy 4d ${\cal N}=1$ supergravity arising from type IIB compactified on a Calabi-Yau orientifold with D7-branes.

In section~\ref{T_stab}, we will review the method of `K\"ahler uplifting' and analytically derive the existence of the meta-stable dS vacuum for the volume modulus of a one-parameter Calabi-Yau compactification with $h^{1,1}=1$ K\"ahler modulus, and then extend this to the case of several K\"ahler moduli $h^{1,1}>1$ explicitly. The interplay of perturbative and non-perturbative effects implies for $h^{1,1}=1$ that here a structure of \emph{two} terms with alternating signs is sufficient to approximate the volume modulus scalar potential and its tunable dS vacuum. This contrasts with the `3-term structure' generically necessary in purely perturbatively stabilized situations~\cite{Saltman:2004jh,Silverstein:2007ac}. For $h^{1,1}>1$ a `3-term structure' reappears for the additional $h^{1,1}-1$ blow-up K\"ahler moduli of a `swiss cheese' Calabi-Yau. 

Finally, we will show that we can express the existence of the meta-stable dS vacuum for the volume modulus in terms of a \emph{sufficient} condition on the microscopic parameters. These are consisting of the fluxes, the D7-brane configuration, and the Euler number of the Calabi-Yau governing the perturbative ${\cal O}(\alpha'^3)$-correction, which are all in turn determined by the underlying F-theory compactification on an elliptically fibred Calabi-Yau fourfold. Thus, the result amounts to a sufficient condition for the existence of meta-stable dS vacua in terms of purely F-theory geometric and topological data which can be satisfied for a sizable subclass of all 4d ${\cal N}=1$ F-theory compactifications, instead of just single `lamp post' models. We also check that our sufficient condition satisfies the necessary condition for meta-stable dS vacua in 4d ${\cal N}=1$ supergravity given in~\cite{Covi:2008ea} and the longevity of the metastable vacuum under tunneling.

Section~\ref{TS_stab} includes the dilaton into a full analytical treatment of the combined dS minimum. We show that supersymmetry breaking happens predominantly in the volume modulus direction, and explicitly determine the shift of the dilaton away from its flux-stabilized supersymmetric locus as suppressed by inverse powers of the volume of the Calabi-Yau.

Section~\ref{TSU_stab_gen} extends the analysis by including an arbitrary number of complex structure moduli with unspecified dependence in the K\"ahler and superpotential. We then show that the shift of the complex structure moduli and the dilaton in general is suppressed by inverse powers of the volume, and that the dilaton and all complex structure moduli generically are fixed at positive-definite masses. Finally, we estimate the backreaction of the shifted dilaton and complex structure moduli onto the volume modulus. The ensuing shift of the stabilized volume is generically found to be small and suppressed by inverse powers of the volume. This crucially extends the sufficient condition for the existence of dS vacua in type IIB F-theory compactifications to a large class of `swiss cheese' style fluxed Calabi-Yau compactifications with arbitrary $h^{1,1}<h^{2,1}$.

In section~\ref{TSU_stab}, we apply our methods to a simple toy model where the K\"ahler and superpotential of complex structure moduli are approximated by the structure found in a torus compactification. We verify the general results of the previous sections, and show that the shifts of the moduli and the backreaction effects are either independent of the number of complex structure moduli $h^{2,1}$, or decreasing as an inverse power of $h^{2,1}$. We conclude in section~\ref{Concl}.

While this paper was being finished, we became aware of~\cite{deAlwis:2011dp}, whose section 2 contains overlapping results with our section~\ref{T_stab}. The main results of section~\ref{T_stab} and~\ref{TS_stab} here have first been presented in talk by one of the authors in~\cite{SFBmeeting}. Additionally, we find numerical disagreement concerning the values of $x$ in section~\ref{T_stab} permissible for a meta-stable dS vacuum of $T$ compared to the results for the same quantity given in section 2 of~\cite{deAlwis:2011dp} due to an approximation used between eq.s (16) and (17) \emph{ibid}.

\section{`K\"ahler uplifting' -- a meta-stable dS vacuum for the K\"ahler modulus} \label{T_stab}

We will start with reviewing the structure of 'K\"ahler uplifted' dS vacua in type IIB flux compactifications on an orientifolded CY threefold~\cite{Westphal:2006tn}. We will at first restrict ourselves to one-parameter models with $h^{1,1}=1$ and $h^{2,1}>1$ so that the Euler number $\chi = 2 (h^{1,1} - h^{2,1}) < 0$ (which will be shown to be part of the the sufficient condition for the existence dS vacua). Later, we will extend the analysis given here to all so-called swiss-cheese Calabi-Yau threefolds with arbitrary $h^{1,1}>1$ and $h^{2,1}>h^{1,1}$, giving a strong indication that the mechanism discussed here works for all threefolds with $\chi <0 $.

For type IIB compactifications on Calabi-Yau orientifolds with 3-form fluxes and D7-branes the effective 4d ${\cal N}=1$ supergravity of the moduli sector is determined by~\cite{Gukov:1999ya,Giddings:2001yu,Grimm:2004uq,Becker:2002nn}

\begin{align}\label{ModelDef}
K &= - 2 \ln \left( \hat{\mathcal{V}} + \alpha'^3 \frac{\hat\xi}{2} \right) - \ln (S+\bar{S}) - \ln \left( - i \int_{CY_3} \bar{\Omega} \wedge \Omega \right) \,\,,\\
W &= W_{0} + \sum_i A_i e^{- a_i T_i}, \quad \text{with } W_0 = \frac{1}{2 \pi} \int_{CY_3} G_{(3)} \wedge \Omega \,\,.
\end{align}

\noindent Note, that this 4d ${\cal N}=1$ supergravity has three branches of vacua. Firstly, we may look for vacua where $|W_0|\ll1$ is tuned small. Then supersymmetric solutions $D_I W=0$ (with $I$ running over all $h^{1,1}$ K\"ahler moduli, $h^{2,1}$ complex structure moduli, and the dilaton $S$) stabilizing all moduli, with 4-cycle volumes ${\rm Re}\,T_i>>1$, are possible including the $\alpha'$-correction discussed above~\cite{Kachru:2003aw}. On swiss-cheese style Calabi-Yau manifolds, a second branch of solutions are the SUSY-breaking AdS vacua of the Large-Volume-Scenario which work for arbitrary $W_0$~\cite{Balasubramanian:2005zx}, and the third branch consists of the `K\"ahler uplifted' solutions studied below, where typically $|W_0|-{\cal O}(1\ldots 10)$ to get dS vacua.

\noindent For one-parameter models we have $\hat{\mathcal{V}} = \gamma (T + \bar{T})^{3/2}$ and we set $\alpha' := 1$. Here 

\begin{align}
\gamma&=\sqrt{3}/(2\sqrt{\kappa}) \quad,\\
\hat\xi&= - \frac{\zeta(3)}{4 \,\sqrt{2}\, (2 \pi)^3} \, \chi \, (S+\bar{S})^{3/2}\quad,
\end{align}

\noindent and $\kappa$ denotes the self-intersection number of the single K\"ahler modulus $T$ in terms of the Poincare-dual 2-cycle volume modulus $v$ of the underlying ${\cal N}=2$ theory prior to orientifolding. The volume of 1-parameter CY threefolds is then given by~\cite{Grimm:2004uq}

\begin{equation}
\hat{\cal V}=\frac{\kappa}{6}\,v^3\equiv\gamma\, (T+\bar T)^{3/2} \quad,\quad \text{Re}\,T = \frac13\,\partial_v \Vol\quad.
\end{equation}

\noindent The flux-superpotential $W_0$ is determined by the integral over the holomorphic 3-form $\Omega$ of the Calabi-Yau and the 3-form flux $G_{(3)}$~\cite{Gukov:1999ya}. The K\"ahler potential $K$ and superpotential $W$ determine the $F$-term scalar potential to be

\begin{equation}
V = e^{K} \left( K^{a\bar{b}} D_a W \overline{D_b W} - 3 |W|^2 \right)
\end{equation}
with $D_a W = W_a + K_a W$, and $a$ runs over the dilaton $S$, the single K\"ahler modulus $T$ and the $h^{2,1}$ complex structure moduli $U_i$. We will now stabilize the K\"ahler modulus 
\begin{equation} T=t+i \tau\quad,
\end{equation}
($\tau$ denotes its axion) using the interplay between the leading perturbative $\alpha'$ correction $\hat\xi$ to the K\"ahler potential~\cite{Becker:2002nn} and non-perturbative corrections to the superpotential. For now, we assume the dilaton $S$ and the complex structure moduli $U_i$ to be stabilized already. Thus, we have to find local stable minima of the scalar potential descending from eq.s~\eqref{ModelDef} assuming $D_SW=D_{U_i}W=0$.


%

Following \cite{Becker:2002nn,Balasubramanian:2004uy,Westphal:2006tn} we can write the resulting scalar potential in the following form
\begin{align}
V(T) &= e^{K} \left( K^{T\bar{T}} D_T W \overline{D_T W} - 3 |W|^2 \right)\\
&= e^{K} \left( K^{T\bar{T}} \left[ W_T \overline{W_T} + (W_T \cdot \overline{W K_T} + c.c) \right] + 3 \hat\xi \frac{\hat{\xi}^2+7\hat{\xi}\hat{\mathcal{V}}+\hat{\mathcal{V}}^2}{(\hat{\mathcal{V}}-\hat{\xi})(\hat{\xi}+2\hat{\mathcal{V}})^2} |W|^2 \right) \,\,.\notag
\end{align}

\noindent Here $K^{T\bar T}$ denotes the $T\bar T$-component of the inverse of the K\"ahler metric $(K_{I\bar J})^{-1}$ where $I,J$ run over all fields involved.

The non-trivial task is to find stationary points of $V(T)$ with respect to $t$. It is straightforward to show that the axionic direction has an actual minimum at $\tau = 0$. The K\"ahler potential does not depend on $\tau$ and the exponential in eq.~\eqref{ModelDef} introduces trigonometric functions $\sin(a\tau)$ and $\cos(a\tau)$ into $V(T)$. Then it can be shown that $V_\tau=0$ for $\tau = n \pi / a$ for $n \in \mathbb{Z}$. We restrict to the case $\tau=0$ so that after insertion of $W_T$ we obtain

\begin{equation}
V(t) = e^{K} \left( K^{T\bar{T}} \left[ a^2 A^2 e^{- 2 a t} + (- a A e^{-a t} \overline{W K_T} + c.c) \right] + 3 \hat\xi \frac{\hat{\xi}^2+7\hat{\xi}\hat{\mathcal{V}}+\hat{\mathcal{V}}^2}{(\hat{\mathcal{V}}-\hat{\xi})(\hat{\xi}+2\hat{\mathcal{V}})^2} |W|^2 \right)\,\,.
\label{VFtWTinserted}
\end{equation}

\subsection{Approximating the scalar potential $V(T)$ in the large volume limit} \label{VFtapprox}

In \cite{Westphal:2006tn}, it was shown that one can get de Sitter minima for $T$ at parametrically large volume $\Vol \simeq \order(100\dots1000)$ and weak string coupling $g_S \simeq 0.1$. The stable minimum is realized at $\hat\xi/(2\Vol) \simeq 0.01$ so small that neglecting higher orders in the $\alpha'$ expansion is well justified and string loop effects are double-suppressed due to the smallness of $g_S$ and the extended no-scale structure~\cite{Cicoli:2007xp}. This minimum can be constructed under the following conditions

\begin{itemize}
 \item Put a stack of $N\simeq \order(30\dots100)$ D7-branes on the single 4-cycle that undergoes gaugino condensation.\footnote{For example, the 2-parameter model $\mathbb{P}_{11169}^4$ was shown in~\cite{Denef:2004dm} to have an F-theory lift containing an $E_8$ ADE-singularity for the condensing gauge group, giving a rank of 30. In general, the achievable rank of the gauge groups is limited for compact CY fourfolds, due to the compactness interfering with enforcing an ADE-singularity of arbitrarily high rank along a given divisor. Still, on compact F-theory fourfolds very large gauge groups with very large ranks can be generated, e.g. in~\cite{Candelas:1997pq} F-theory was compactified to 4d on a compact fourfold to yield a gauge group with 251 simple factors, the largest of which was $SO(7232)$.} The parameter $A$ is assumed to be $\order(1)$.
 \item Choose the flux induced superpotential $W_0 \simeq \order(-30)$ and the parameter $\hat\xi \simeq \order(10)$. Note that a $W_0$ of this rather large magnitude does not induce problematic back reactions, as in type IIB the fluxes are imaginary self-dual (ISD) and of (1,2) or (0,3) type which limitates the back reaction to the warp factor.
\end{itemize}

In this setup, one typically obtains a minimum at $T \simeq \order(40)$ so that the non-perturbative contribution to the superpotential $A e^{-a T}$ is small enough to also trust the Ansatz for the non-perturbative superpotential.

We now want to give a parametric understanding of this scenario by approximating the scalar potential eq.~\eqref{VFtWTinserted} under the constraint of the typical values of the parameters ${a,A,W_0,\hat\xi,\gamma}$. We use the condition $\hat\xi/(2\Vol) \simeq 0.01$ and the validity of the non-perturbative superpotential:

\begin{equation}
\hat{\mathcal{V}} \gg \hat{\xi}, \qquad |W_0| \gg A e^{- a t}\,\,.
\label{approxconditions}
\end{equation}

Under these approximations, the K\"ahler Potential and its derivatives simplify in the following way:

\begin{align}
K &= - 2 \ln \left( \hat{\mathcal{V}} + \frac{\hat\xi}{2} \right) \simeq - 2 \ln \left( \hat{\mathcal{V}} \right) \,\,,\notag\\
K_T &= \frac{-3 \gamma^{2/3} \sqrt[3]{\hat{\mathcal{V}}}}{\hat{\mathcal{V}}+\frac{\hat\xi}{2}} \simeq \frac{-3 \gamma^{2/3}}{\hat{\mathcal{V}}^{2/3}} \,\,,\notag\\
\left(K_{T\bar{T}}\right)^{-1} &= \gamma^{-4/3} \frac{\sqrt[3]{\hat{\mathcal{V}}}(4\hat{\mathcal{V}}^2+\hat\xi \hat{\mathcal{V}}+4\hat{\xi}^2)}{12(\hat{\mathcal{V}}-\hat\xi)} \simeq \frac{\hat{\mathcal{V}}^{4/3}}{3\gamma^{4/3}}\,\,.
\end{align}
Also the last term of eq.~\eqref{VFtWTinserted} simplifies under the approximation eq.~\eqref{approxconditions}. Implementing eq.~\eqref{approxconditions}, the scalar potential eq.~\eqref{VFtWTinserted} becomes

\begin{equation}
V(t) \simeq \frac{e^{-2 a t} (3 a A^2 + a^2 A^2 t)}{
 6 \gamma^2 t^{2}} + \frac{a A e^{-a t} W_0}{2 \gamma^2 t^2} + \frac{3 W_0^2 \hat\xi}{
 64 \sqrt{2} \gamma^3 t^{9/2}}\,\,.
 \label{VFt3term}
\end{equation}
We also neglect the term $\propto e^{-2 a t}$ since it is suppressed by one more power of $e^{- a t}$ compared to the second term in eq.~\eqref{VFt3term} and obtain a `2-term structure' for the scalar potential

\begin{equation}
V(t) \simeq \frac{a A e^{-a t} W_0}{2 \gamma^2 t^2} + \frac{3 W_0^2 \hat\xi}{
 64 \sqrt{2} \gamma^3 t^{9/2}}\,\,.
\label{VFt2termt}
\end{equation}
Note that the flux-superpotential is negative, $W_0 < 0$, so that the \emph{two} terms have opposite sign and a minimum is in principle allowed. Eq.~\eqref{VFt2termt} is a drastic simplification of the rather complicated scalar potential eq.~\eqref{VFtWTinserted} that allows us to extract an analytic condition on the parameters to obtain a meta-stable de Sitter vacuum. Factorizing eq.~\eqref{VFt2termt}, we can write it in terms of two characteristic variables $x=a \cdot t$ and $C$

\begin{equation}
V(x) \simeq \frac{- W_0 a^3 A}{2 \gamma^2} \left( \frac{2 C}{9 x^{9/2}} - \frac{e^{-x}}{x^2} \right), \quad C = \frac{-27 W_0 \hat\xi a^{3/2}}{64 \sqrt{2} \gamma A}\,\,.
\label{VFt2termx}
\end{equation}
The overall constant in eq.~\eqref{VFt2termx} does not influence the extrema of this potential. For completeness, we mention that the stationary point in the axionic direction $\tau=0$ is always a minimum since the mass

\begin{equation}\label{mtau0}
 V_{\tau\tau} = -\frac{a^3 A e^{-a t} W_0}{2 \gamma^2 t^2}>0\quad{\rm if}\quad W_0<0\;\;.
\end{equation}
The mass matrix $V_{ij}$ for $i,j \in \{t,\tau\}$ is diagonal since the mixed derivative $V_{t\tau}$ vanishes at $\tau = 0$.

Note, that it is the presence of the exponential factor in the negative term with the slower inverse power-law dependence on $x$, which renders this term as a `negative middle term' in terms of the analysis of~\cite{Silverstein:2007ac}. Here, however, this term shuts down exponentially fast for large enough $x$. This combined behavior of being a power-law at small $x$ and an exponential at larger $x$ is responsible for the fact, that a `2-term' combination with a single positive inverse power-law term is enough to obtain a tunable dS vacuum.

\subsection{A sufficient condition for meta-stable de Sitter vacua} \label{dScond}

To calculate extrema of eq.~\eqref{VFt2termx} we need to calculate the first and second derivative with respect to $x$ $(V' = \frac{\partial V}{\partial x})$

\begin{align}
V'(x) & =  \frac{- W_0 a^3 A}{2 \gamma^2} \frac{1}{x^{11/2}} \left( C - x^{5/2} (x+2) e^{-x} \right)\,\,,\\
V''(x) &= \frac{- W_0 a^3 A}{2 \gamma^2} \frac{1}{x^{13/2}} \left( \frac{11}{2} C - x^{5/2} (x^2+4 x +6) e^{-x}\right)\,\,.\label{VF''x}
\end{align}
Solving for an extremum $V'(x) = 0$ yields

\begin{equation}
x^{5/2} (x+2) e^{-x} = C
\label{xmintrans}
\end{equation}
which cannot be solved explicitly in an analytic way. Plotting the approximate expression eq.~\eqref{VFt2termx} of $V(x)$ for different values of the constant $C$ in figure $\ref{fig_CondOnC}$ we observe the following behavior:

\begin{figure}[t!]
\centering
\includegraphics[width=\linewidth]{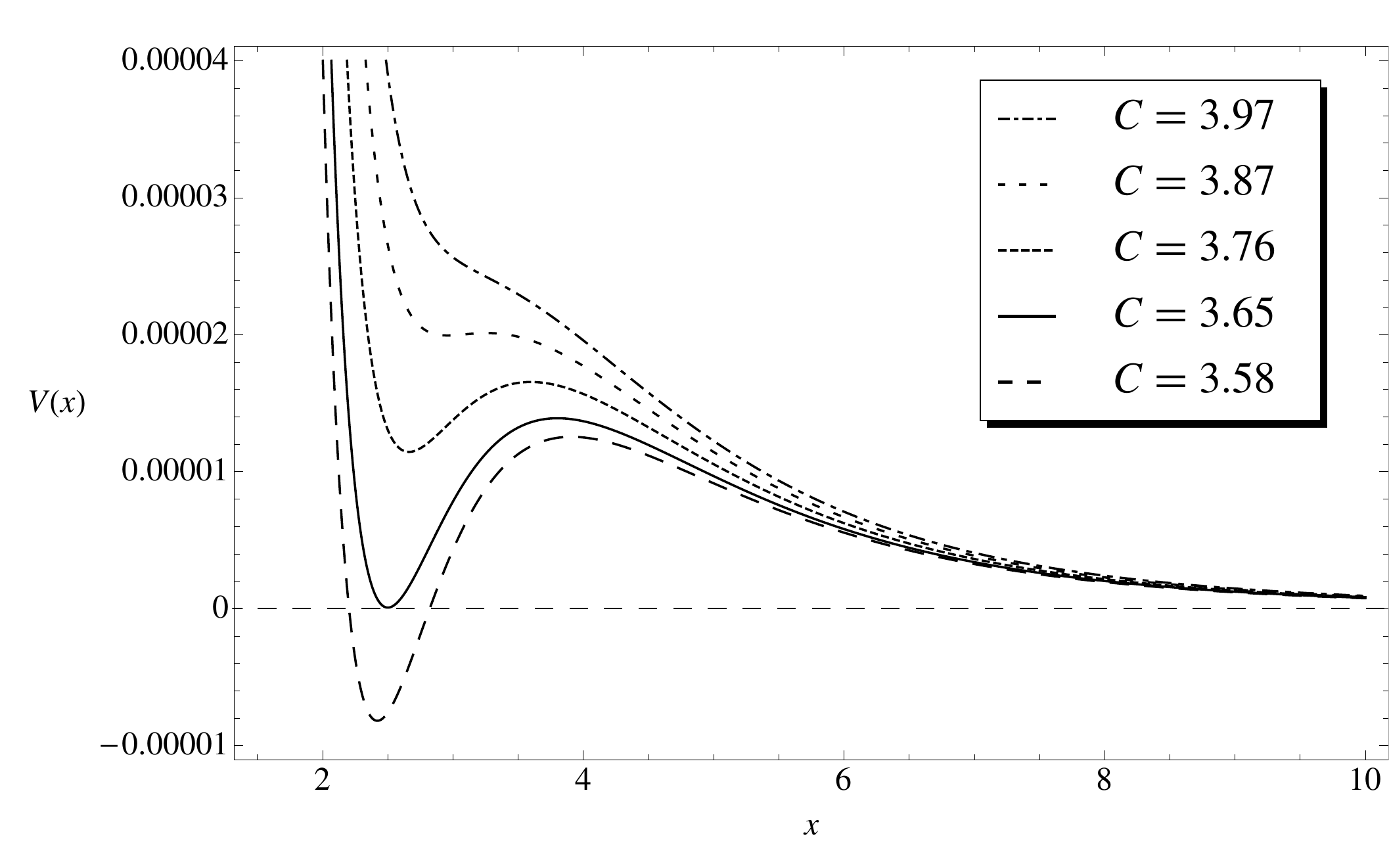}
\caption{The approximate 2-term scalar potential $V(x)$ from eq.~\eqref{VFt2termx} for different values of $C$.}
\label{fig_CondOnC}
\end{figure}

We see that with growing $C$ we first obtain an AdS minimum. This minimum breaks supersymmetry since

\begin{equation}
 F_T \simeq \frac{-3 W_0}{2 t \Vol} \neq 0\,\,. 
\end{equation}
Then at some point the minimum transits to dS, and for even larger $C$ the potential eventually develops a runaway in the $x$ direction. We can analytically calculate the window for $C$ where we obtain a meta-stable de Sitter vacuum by identifying:

\begin{itemize}
 \item Lower bound on $C$: $V(x_{min}) = V'(x_{min}) = 0$
 \item Upper bound on $C$: $V'(x_{min}) = V''(x_{min}) = 0$
\end{itemize}

In both cases we have to solve two equations for two variables $\{x_{min},C\}$. For instance, one can use eq.~\eqref{xmintrans} to replace $C e^x$ in $V(x)=0$ for the lower and in $V''(x)=0$ for the upper bound which gives equations maximally quadratic in $x$ and then use eq.~\eqref{xmintrans} again to calculate $C$. In both cases, there exists only one solution with $x_{min} > 0$.

\begin{itemize}
 \item Lower bound on $C$:\\ $\{x_{min},C\} = \{\frac{5}{2},\frac{225}{8} \sqrt{\frac{5}{2}} e^{-\frac{5}{2}}\} \simeq \{2.5,3.65\}$
  \item Upper bound on $C$:\\ $\{x_{min},C\} \simeq \{\frac{3 + \sqrt{89}}{4} ,3.89\} \simeq \{3.11,3.89\}$
\end{itemize}
The region close to $\{x_{min},C\}$ is the one relevant for obtaining a small positive cosmological constant suitable for describing the late-time accelerated expansion of the universe. For $a = 2\pi/100$ the lower bound on $x$ corresponds to a volume $\Vol \simeq 100$ so we are indeed at parametrically large volume. 
The allowed window for $C$ to obtain meta-stable de Sitter vacuum is approximately

\begin{equation}
\boxed{3.65 \lesssim \frac{-27 W_0 \hat\xi a^{3/2}}{64 \sqrt{2} \gamma A} \lesssim 3.89}
\label{dSCondeq}
\end{equation}

In sections $\ref{TS_stab}$ and $\ref{TSU_stab_gen}$, we will show that fulfilling condition eq.~\eqref{dSCondeq} is still sufficient to obtain a meta-stable minimum of the scalar potential when all the remaining moduli fields of the Calabi-Yau, i.e. the dilaton and the complex structure moduli, are included in the stabilization analysis. Hence, this is truly a sufficient condition for meta-stable de Sitter vacua and no tachyonic instabilities occur by including further moduli, contrary to the standard KKLT scenario \cite{Choi:2004sx,Lust:2005dy}.

\subsection{$h^{1,1}>1$}

We will now proceed to show explicitly that the above argument can be extended to the full class of all Calabi-Yau threefolds with $h^{1,1}>1$ arbitrary and $\chi<0$ which are of `swiss cheese' type\footnote{For the two K\"ahler moduli of $\mathbb{P}^4_{[1,1,1,6,9]}$ the K\"ahler uplifted dS minimum was found numerically first in~\cite{AbdusSalam:2007pm}.}. A `swiss cheese' type Calabi-Yau is characterized by a classical volume given by
\begin{equation}
\Vol=\sum_{I=1}^{h^{1,1}}\frac16\kappa_{III}\, (v^I)^3=\gamma\, (T+\bar T)^{3/2}-\sum_{i=2}^{h^{1,1}}\gamma_i\, (T_i+\bar T_i)^{3/2}
\end{equation}
where $v^I$ is the 2-cycle modulus, $\kappa\equiv \kappa_{111}>0$, $\kappa_{iii}<0$ for $i=2\ldots h^{1,1}$ and
\begin{eqnarray}
t\equiv{\rm Re}\, T&=&\frac13 \partial_v \Vol \quad{\rm and}\quad \gamma=\frac{\sqrt 3}{2\sqrt{\kappa}} \nonumber\\
&&\\
t_i\equiv{\rm Re}\,T_i&=& \frac13\partial_{v^i}\Vol \quad{\rm and}\quad \gamma_i=\frac{\sqrt 3}{2\sqrt{-\kappa_{iii}}}\;\;\forall\;i=2\ldots h^{1,1}\quad.\nonumber
\end{eqnarray}
This structure allows us to invert and get
\begin{equation}\label{vIoftI}
v_I=2\sqrt 2 \gamma_I \sqrt{t_I}\quad.
\end{equation}
Thus the classical volume of such Calabi-Yaus has a $(+---\ldots)$ signature in intersection number space. We will look for dS vacua which satisfy ${\rm Re}\,T_i\ll {\rm Re}\,T$ for $i=2\ldots h^{1,1}$ such that $\Vol\sim \gamma\, (T+\bar T)^{3/2}$, such the $h^{1,1}-1$ blow-up K\"ahler moduli form the `holes' of the `swiss cheese'. This entails choosing the $a_i$ for $i=2\ldots h^{1,1}$ of the nonperturbative superpotential effects on the associated 4-cycles such that $a_i\gg a\equiv a_1$ while enforcing $a_i t_i >1$ to maintain the validity of the one-instanton approximation.

We will again determine the leading terms in $\hat\xi/\Vol$ as before. The scalar potential reads
\begin{eqnarray}
V &=& e^{K} \bigg( K^{T_I\bar{T}_J} \left[ a_I a_J A_I A_J e^{- a (T_I+\bar T_J)} + (- a_I A_I e^{-a T_I} \overline{W K_{T_J}} + c.c) \right]\nonumber\\
&&\qquad  + 3 \hat\xi \frac{\hat{\xi}^2+7\hat{\xi}\hat{\mathcal{V}}+\hat{\mathcal{V}}^2}{(\hat{\mathcal{V}}-\hat{\xi})(\hat{\xi}+2\hat{\mathcal{V}})^2} |W|^2 \bigg)\quad.
\end{eqnarray}
Guided by eq.~\eqref{VFt2termt}, we extract the terms linear in $\hat\xi$ and in $e^{-a T_I}$, suppressing terms which are of order $e^{-a (T+\bar T_J)}$ or $\hat\xi e^{-a T_I}$ as they are subleading in the limit we are considering. We will see that we are forced to keep terms of order $e^{-a (T_i+\bar T_j)}$ as these will turn out to be of a relevant order in $\hat\xi/\Vol$ once the condition of the minimum is imposed.  Using the fact that $K_{T_I}=K_{\bar T_I}$ for our choice of $\Vol$, the relevant part of the potential thus reads
\begin{eqnarray}\label{LeadingOrderPot}
V&=&-\frac{W_0}{\Vol^2}K^{T_I\bar T_J}\left(a_I A_I e^{-a T_I}K_{\bar T_J}+c.c\right)+\frac{3\hat\xi W_0^2}{4\Vol^3}\quad.
\end{eqnarray}
We have
\begin{equation}\label{KahlerDeriv}
K_T=K_{\bar T}=-\frac{3\sqrt 2\gamma \sqrt t}{\Vol}+{\cal O}\left(\frac{\hat\xi}{\Vol}\right)\quad,\quad K_{T_i}=K_{\bar T_i}=\frac{3\sqrt 2 \gamma_i \sqrt{t_i}}{\Vol}+{\cal O}\left(\frac{\hat\xi}{\Vol}\right)
\end{equation}
and the inverse K\"ahler metric can be found~\cite{Bobkov:2004cy} to be
\begin{equation}
K^{T_I\bar T_J}=-\frac29 (2\Vol +\hat\xi)\kappa_{IJK}v^K+\frac{4\Vol -\hat\xi}{\Vol-\hat\xi} t_I t_J\quad.
\end{equation}
Now we can apply our limit $\hat\xi/\Vol\ll 1$, use that the $\kappa_{IJK}=\kappa_{III}$ are diagonal, and implement that $t_i\ll t$ for $i=2\ldots h^{1,1}$. We then find using eq.~\eqref{vIoftI} that
\begin{eqnarray}\label{KahlermetricInv}
K^{T\bar T}&=&-\frac49\Vol \kappa v+4t^2=-\frac49 \Vol\frac{3}{4\gamma^2}2\sqrt 2\gamma \sqrt t+4t^2\simeq\frac{\sqrt 2}{3}\Vol\frac{\sqrt t}{\gamma}\nonumber\\
&&\nonumber\\
K^{T\bar T_i}&=& 4 \,t_i t \\
&& \nonumber\\
K^{T_i\bar T_i}&=& -\frac49\Vol \kappa_{iii} v^i+4t_i^2=-\frac49 \Vol\frac{-3}{4\gamma_i^2}2\sqrt 2\gamma_i \sqrt{t_i}+{\cal O}(1)\simeq\frac{2\sqrt 2}{3}\Vol\frac{\sqrt{t_i}}{\gamma_i}\nonumber
\end{eqnarray}
while $K^{T_i\bar T_j}=4 t_i t_j\ll {\cal O}(v^I \Vol)$ can be dropped in this limit. Plugging eq.s~\eqref{KahlerDeriv} and \eqref{KahlermetricInv} into 
the potential eq.~\eqref{LeadingOrderPot}, we get
\begin{eqnarray}\label{LeadingOrderPot2}
V&=&\frac{4W_0}{\Vol^2}\left(a t A e^{-a t} \cos(a\tau)+ \sum_{i=2}^{h^{1,1}}a_i t_i A_i e^{-a_i t_i}\cos(a_i\tau_i)\right)+\frac{3\hat\xi W_0^2}{4\Vol^3}\nonumber\\
&&\qquad +\sum_{i=2}^{h^{1,1}}\frac43 \frac{a_i A_i^2}{\Vol^2}\,\frac{\sqrt{t_i}}{\gamma_i}e^{-2 a_i t_i}\left[\frac{\Vol}{\sqrt 2}+3\gamma_i\sqrt{t_i} (a_i t_i+1)\right]\quad.
\end{eqnarray}
The cross terms $\sim K^{T\bar T_i}$ are relevant to obtain the correct sign of the axion terms in the first round bracket. The terms $\sim e^{-2a_i t_i}$ look subleading. However, at the prospective minimum one can show that $e^{-a t}\sim e^{-a_i t_i}\sim \hat\xi/\Vol$. This implies, that the terms $\sim e^{-2a_i t_i}$ are in fact $\sim \hat\xi^2/\Vol^3$ (including the factor of $\Vol$ in the rectangular bracket) and are thus of relevant order for minimization.

This potential has a full minimum at $at\simeq a_it_i\simeq 3$ for $i=2\ldots h^{1,1}$, and $\tau=\tau_i=0$, if the quantity $C$ defined in eq.~\eqref{VFt2termx} satisfies a structurally similar bound on $C$ as in the one-parameter case discussed above. However, the numerical interval of $C$-values allowed by the metastability conditions increases slowly with $h^{1,1}$. The size of $C \sim |W_0|$ for given $\hat\xi$, intersection numbers, and gauge group ranks. As the maximum size of $|W_0|$ is given by the maximum available fluxes, this implies an upper bound on $h^{1,1}$ as the flux is limited by the tadpole constraint quantified by the Euler characteristic of the F-theory elliptic fourfold.  The magnitude of $\chi_{CY_4}$ can be easily as large as ${\cal O}(10^4)$, so this is not a particularly strong bound.

Thus we expect this minimum to persist for all `swiss cheese' Calabi-Yau threefolds of arbitrary $h^{1,1}>1$. Moreover, the way how the additional K\"ahler moduli enter the leading terms of the scalar potential implies that the inclusion of dilaton and complex structure stabilization discussed in the subsequent chapters will also extend to the $h^{1,1}>1$ case by virtue of its viability for the first K\"ahler modulus. Finally, as the quantity $\kappa_{IJK}v^K$ is a matrix with signature $(1, h^{1,1}-1)$ (one plus and the rest minus)~\cite{Candelas:1990pi}, we expect the overall sign structure of eq.s~\eqref{KahlerDeriv} and \eqref{KahlermetricInv} to persist even for general non-`swiss cheese' type Calabi-Yau threefolds. As this will lead to a potential with the same basic structure as eq.~\eqref{LeadingOrderPot2} we may expect this mechanism of stabilizing all K\"ahler moduli directly into a dS vacuum via `K\"ahler uplifting' to extend to all Calabi-Yau threefolds with $\chi <0$.

\subsection{Suppressing flux-induced $\alpha'$-corrections}

We now want to discuss potentially dangerous flux-induced $\alpha'$-corrections originating from $|W_0| \simeq \order{(30)}$. The leading $\alpha'$ correction we used to stabilize the volume modulus in section~\ref{T_stab} to the K\"ahler potential can be derived from the $\mathcal{R}^4$ term in the 10d effective supergravity action, where $\mathcal{R}$ is the 10d Ricci scalar. However, for a large flux-induced superpotential, corrections to the scalar potential descending from the $\mathcal{R}^3 G_{(3)}^2$ term in the 10d effective action might become relevant even though suppressed by higher powers of the inverse volume~\cite{Conlon:2005ki} in the scalar potential. To trust our analysis of dS vacua in section~\ref{T_stab} - which takes only the $\mathcal{R}^4$ $\alpha'$-correction into account - we need to ensure
\begin{equation}
\Delta V_{\mathcal{R}^3 G_{(3)}^2} \sim \frac{|G_{(3)}|^2}{\Vol^{11/3}} \sim \frac{W_0^2}{\Vol^{11/3}} < \Delta V_{\mathcal{R}^4} \sim \frac{\hat\xi}{\Vol^3} \quad, \label{fluxinducedsmaller}
\end{equation}
where we have used that $|W_0|\sim {\cal O}(|G_{(3)}|)$ barring fortuitous cancellations.
For the `K\"ahler uplifting' regime $\Vol \simeq \order(100\dots1000)$, $\hat\xi \simeq \order(10)$ and $|W_0| \simeq \order(30)$ the inequality~\eqref{fluxinducedsmaller} is not a priori fulfilled but rather $\Delta V_{\mathcal{R}^3 G_{(3)}^2} \simeq \Delta V_{\mathcal{R}^4}$, promoting flux-induced $\alpha'$-corrections problematic.

We can ensure the desired hierarchy between the corrections by demanding the choices of all explicit flux quanta to be of ${\cal O}(1)$. In general, this implies that $|W_0| \simeq \order(1)$ which is effectively a percent-order fine-tuning. Since $G_{(3)} = F_{(3)} - S\cdot H_{(3)}$, we can write
\begin{equation}
W_0 = \frac{1}{2 \pi} \int_{CY_3} F_{(3)} \wedge \Omega - S \, \frac{1}{2 \pi} \int_{CY_3} H_{(3)} \wedge \Omega \equiv C_1 - C_2\cdot S\quad.
\end{equation}
We will show in section~\ref{TS_stab}, that the dilaton $S$ will be approximately stabilized supersymmetrically at $\text{Re}(S) = g_S^{-1} = - C_1/C_2$, so that $\langle W_0\rangle = 2 \,C_1$. We can reduce $|W_0|$ from $\order(30)$ to $\order(1)$ and stay in a dS vacuum if we also reduce $g_S$ because according to the sufficient condition for dS vacua in eq.~\eqref{dSCondeq}
\begin{equation}
 W_0 \hat\xi \propto \frac{W_0}{g_S^{3/2}} = \text{const.}
\end{equation}
Note that reducing $g_S$ also improves the approximation of neglecting string loop effects. This translates into the following requirements on the two functions $C_1$ and $C_2$:
\begin{align}
 g_S &= -C_2/C_1 \ll 1\notag\\
 W_0 &= 2\,C_1 \sim \order(1)\label{finetuneC1C2}
\end{align}
As $C_1$ and $C_2$ are functions of the flux-quanta and the VEV's of the complex structure moduli, eq.~\eqref{finetuneC1C2} can be fulfilled by choosing $\order(1)$ flux-quanta for $F_{(3)}$ and $H_{(3)}$ while choosing the VEV's of the complex structure moduli in $C_2$ such that $C_2 \ll 1$. The bound on $C_2$ which depends on the $H_{(3)}$ flux-quanta is more stringent than the bound on $C_1$ which depends on the $F_{(3)}$ flux-quanta. This follows from a look at the superpotential $W_0 = C_1 - S\,C_2$ , and the way how $H_{(3)}$ enters the 10d type IIB bulk action. The crucial point is that the bulk terms $\int d^{10}x\sqrt{-g} e^{-2\phi} H^2_{(3)}$ and $\int d^{10}x\sqrt{-g} e^{-2\phi} R$ enter with the same powers of the string coupling. Therefore, the dominant flux-induced ${\cal O}(\alpha'^3)$-correction which is $\sim R^3 H_{(3)}^2$ will give a correction to the K\"ahler potential which scales with same dilaton dependence as the one from the $R^4$-term, but is suppressed by an additional power $1/\Vol^{2/3}$. This is because $R$ scales as $1/\Vol^{1/3}$ while $H_{(3)}^2\sim 1/\Vol$. By keeping $\int _{\Sigma_3}H_{(3)}$ of ${\cal O}(1)$ we can then suppress the flux-induced $\alpha'$-correction by large volume. Furthermore, since the $F_{(3)}$ flux is suppressed by a further power of $g_s$ the bound is obviously stronger on the $H_{(3)}$ flux. Thus, tuning of $C_1$ and $C_2$ can alleviate the problem of flux-induced $\alpha'$-corrections. Reducing fluxes requires additional contributions to fulfill tadpole constraints. These contributions can be supplied by D3- and magnetized D7-branes.

\subsection{F-theory interpretation}

Eq.~\eqref{dSCondeq} forms a crucial result of our analysis. It represents an explicit condition relating two topological properties of the CY threefold, the self-intersection number $\kappa$ of its volume modulus, and its Euler characteristic $\chi$ (via $\hat\xi=\hat\xi(\chi)$), to the flux superpotential and the rank of condensing D7-brane gauge group. \footnote{This verifies the numerical evidence fount in~\cite{Westphal:2006tn}, indicating that one can trade of larger $W_0$ for smaller $\hat\xi$ and still obtain a de Sitter minimum. This now is obvious from eq.~\eqref{dSCondeq}.}

Let us briefly comment here on the link to F-theory. Type IIB warped flux compactifications on an O7-orientifolded CY threefold with D3- and D7-branes with varying axio-dilaton can be described as the Sen limit of F-theory compactified on an elliptically fibred CY fourfold. The CY threefold then is the double-cover of the base of the elliptic fibration under the orientifold projection in the Sen limit~\cite{Sen:1997gv}. The F-theory description unifies the different objects of the CY threefold, the O7-plane and the D7-branes including the non-abelian gauge theories into the geometry and topology of the elliptically fibred CY fourfold. In particular, points in the base of the fibration, where the torus fibre of the Weierstrass model degenerates via a vanishing 1-cycle, describe D7-branes. In consequence, D7-brane stacks with their non-abelian gauge groups are geometrized into the notion of ADE-type singularities at the points in the base where the torus fibre degenerates. The type IIB 3-form fluxes $H_{(3)}$ and $F_{(3)}$, in turn, descend from a single 4-form flux $G_{(4)}$ on the F-theory fourfold. Finally, the Euler characteristic $\chi$, and $h^{1,1}$, $h^{2,1}$ are completely determined in terms of the topological data of the fourfold.

Using this information, we immediately see that the sufficient condition for the existence of 'K\"ahler uplifted' dS vacua in type IIB becomes particularly elegant criterion in the underlying F-theory construction in so far, as the relation eq.~\eqref{dSCondeq} places a constraint on the geometry and topology of the fourfold and the 4-form flux $G_{(4)}$:

\begin{itemize}
\item The data entering $W_0$,  $\xi$, and $\gamma(\kappa)$, which consists of $\chi$, $h^{2,1}$, the intersection number(s), and the periods of the threefold, are completely determined in terms of the topological data of the fourfold, and the 4-form flux $G_{(4)}$.
\item The rank of the D7-brane gauge group entering $a$ is determined by the ADE singularity enforced at the degeneration point of the Weierstrass model describing the elliptic fibration.
\end{itemize}
Thus,  the sufficient condition eq.~\eqref{dSCondeq} represents a purely geometrical and topological constraint on the fourfold in F-theory except for the constraint on $G_{(4)}$.

\subsection{The necessary curvature condition}

The discussion so far has constituted a sufficient condition for the existence of meta-stable 'K\"ahler uplifted'  dS vacua in type IIB on a CY orientifold. Let us pause here for a moment, and compare this condition to the necessary condition of positive sectional curvature of the K\"ahler potential which~\cite{Covi:2008ea} derived from a general 4d ${\cal N}=1$ supergravity argument. The statement there is that a meta-stable dS vacuum cannot exist unless  the sectional curvature of the full K\"ahler potential of a given model
\begin{equation}
\lambda\equiv 2 g_{i\bar\jmath}G^iG^{\bar\jmath}-R_{i\bar\jmath m\bar n} G^iG^{\bar\jmath}G^mG^{\bar n} > 0
\end{equation}
is positive, where
\begin{equation}
g_{i\bar\jmath}\equiv \partial_i\bar\partial_{\bar\jmath}K,\quad\quad G^i\equiv e^{-G/2}F^i,\quad\quad G=K+\ln |W|^2\,\,,
\end{equation}
and $R_{i\bar\jmath m\bar n}(g_{i\bar\jmath})$ is the Riemann tensor of the scalar manifold. For our case of the leading order ${\cal O}(\alpha'^3)$ correction breaking no-scale and supplying the dominant direction of supersymmetry breaking $F_T$ (this will be shown in the subsequent sections), this condition is equivalent to~\cite{Covi:2008ea}
\begin{equation}\label{NessCond}
\frac{\hat\xi}{8{\cal V}}>\frac{2\,\langle V\rangle}{105\, m_{3/2}^2}\quad.
\end{equation}
We do now see that satisfying the sufficient condition given here implies satisfaction of eq.~\eqref{NessCond}, as 
\begin{equation}
m_{3/2}^2\equiv e^K |\langle W\rangle|^2\simeq e^K |\langle W_0\rangle|^2 > 0
\end{equation}
is guaranteed always in the minimum due to $|\langle W_0\rangle|\sim {\cal O}(1) \gg |e^{-a \langle T\rangle}|$, while tuning $W_0$ allows $\langle V\rangle\simeq 0$ to ${\cal O}(10^{-h^{2,1}})$.

Note that satisfying eq.~\eqref{NessCond} requires $\hat\xi>0$ for true dS vacua, which fixes the sign of $\hat\xi$ and thus $\chi$. This is consistent with the extremum conditions eq.s~\eqref{xmintrans}, \eqref{mtau0} together with definition of $C$ in eq.~\eqref{VFt2termx}, as they too dictate $W_0<0 \Leftrightarrow \hat\xi>0$.

\begin{table}[ht!]
\vskip 1cm
\centering
  \begin{tabular}{c||c|c|c|c}
   & $\;\;\langle t\rangle\;\;$ & $m_t^2$ & $m_\tau^2$ & $m^{2}_{3/2}$\\
  \hline
  exact & $43.0$ & $9.8 \cdot 10^{-4}$ & $2.5 \cdot 10^{-3}$ & $1.3 \cdot 10^{-2}$\\
  approx. & $39.8$ & $1.4 \cdot 10^{-3}$ & $3.4 \cdot 10^{-3}$ & $1.9 \cdot 10^{-2}$\\
  \end{tabular}
  \caption{Numerical results for the minimum $\langle t\rangle$, the moduli masses $m_t^2$, $m_\tau^2$ and the gravitino mass $m^{2}_{3/2}$. The exact results are obtained numerically from eq.~\eqref{VFtWTinserted} for $W_0 = -32.35$, the approximate results from eq.s~\eqref{VFt2termt} and eq.~\eqref{mtau0} for $W_0 = -37.73$. The masses where determined by diagonalization of the Hessian of the relevant scalar potential, and multiplying the eigenvalues with $K^{T\bar T}$ for canonical normalization of the kinetic terms.}
  \label{tab_numT}
\end{table}

We can rewrite eq.~\eqref{NessCond} by inserting the 2-term potential of eq.\eqref{VFt2termx} and the gravitino mass $m_{3/2}^2 \simeq (W_0/\Vol)^2$:
\begin{equation}
 1 > \frac{2}{35} \left(2-\frac{9 e^{-x} x^{5/2}}{C}\right)\quad.
\end{equation}
The only remaining parameters are $x$ and $C$. This allows us to check the necessary curvature condition at the upper limit for $\{x,C\} = \{3.11,3.89\}$, i.e. where the meta-stable dS minimum becomes a saddelpoint in the $t$-direction, see section~\ref{dScond}. We find
\begin{equation}\label{nesscondforuplim}
 1 > \frac{1}{140} \left(9 \sqrt{89} - 83 \right) \simeq 0.014\quad.
\end{equation}
We do not necessarily expect eq.~\eqref{NessCond} to be violated since this is a necessary condition, i.e. a dS vacuum does not have to exist even though the inequality is fulfilled. However, eq.~\eqref{nesscondforuplim} is far from being saturated which suggests that the space of actually meta-stable dS vacua may be significantly smaller than the space of candidate vacua allowed by the necessary condition.

\subsection{Numerical example}

Let us display our above analysis with a numerical example from \cite{Westphal:2006tn}:

\begin{equation}
 a = \frac{2 \pi}{100},\quad W_0 = -32.35,\quad A = 1,\quad \gamma = \frac{\sqrt{3}}{2 \sqrt{5}},\quad \hat\xi = 7.98\,\,.
 \label{exampleT}
\end{equation}
The choice for $\gamma$ is that for the quintic $C\mathbb{P}^4_{1,1,1,1,1}$ which has intersection number $\kappa=5$. The meta-stable minimum of the exact potential eq.~\eqref{VFtWTinserted} lies at $t=43$ so that indeed the approximations in eq.~\eqref{approxconditions} are well justified.

\begin{figure}[ht!]
\centering
\includegraphics[width= \linewidth]{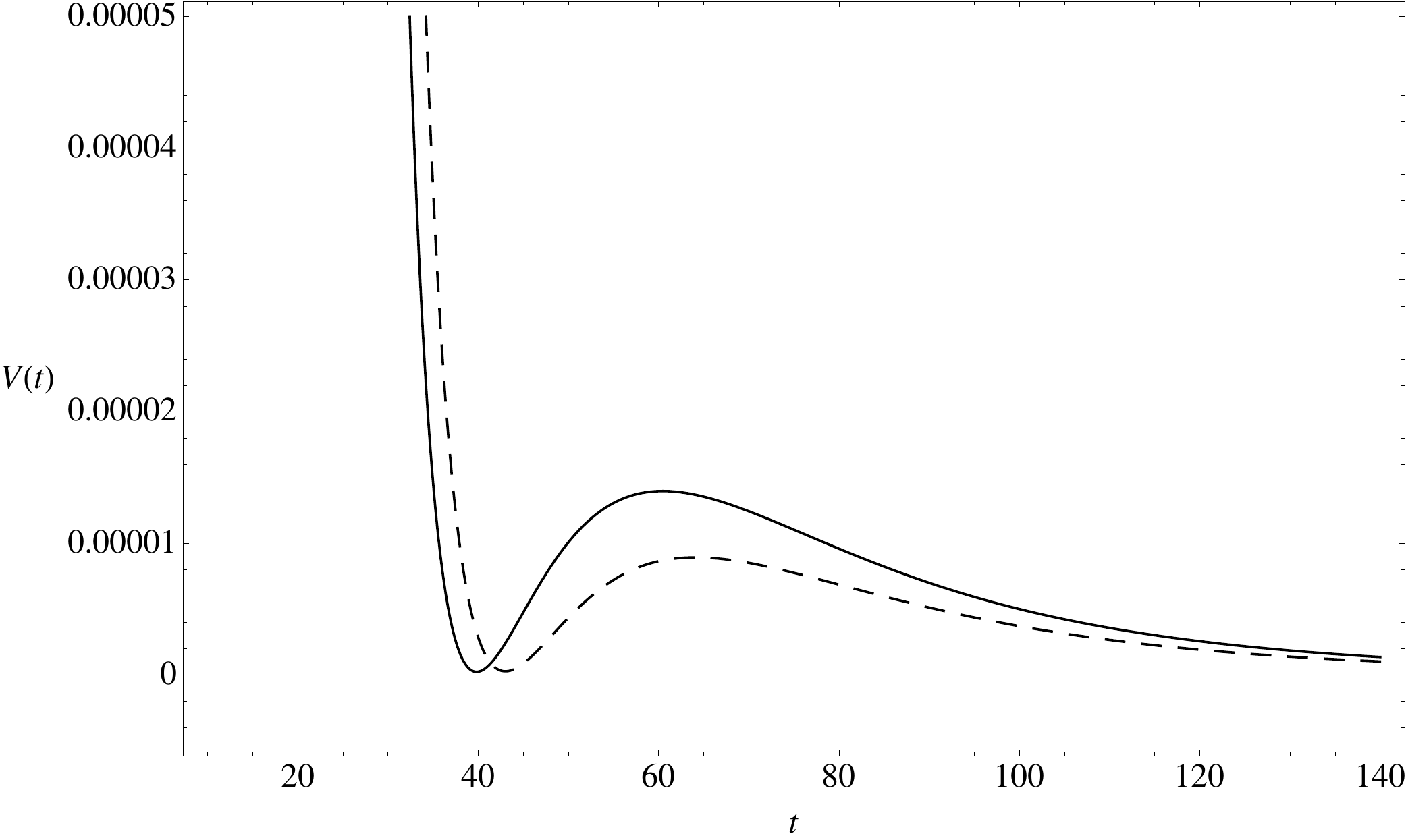}
\caption{Dashed curve: the exact potential with parameters $W_0 = -32.35$. Solid curve: the approximate potential with parameters $W_0 = -37.73$, $C=3.652$.}
\label{fig_VexactVapproxT}
\end{figure}

\begin{equation}
 \frac{\hat\xi}{\Vol} \simeq 0.03 \ll 1, \qquad \frac{A e^{-a t}}{|W_0|} \simeq 0.002 \ll 1 \,\,.
\end{equation}
In figure $\ref{fig_VexactVapproxT}$, we compare the exact potential eq.~\eqref{approxconditions} to the approximate potential eq.~\eqref{VFt2termx} for the parameters eq.~\eqref{exampleT}.

We see that the two curves agree sufficiently to justify the parametric understanding drawn out of the 2-term potential eq.~\eqref{VFt2termx}. The minimum of the approximate potential is located at $t\simeq40$. We give a summary of the numerical results for the moduli VEVs and masses in table $\ref{tab_numT}$.

\subsection{Vacuum decay}\label{decay}
We will now briefly digress to discuss the (meta)stability of the dS vacua we just found. The vacua are local minima of the scalar potential and thus are classically stable. However, they will decay non-perturbatively by a tunneling process. There are two known instanton solutions to the Euclidean equations of motion in the saddle point approximation -- the Coleman-DeLuccia (CdL)~\cite{Coleman:1977py,Callan:1977pt,Coleman:1980aw} and the Hawking-Moss (HM)~\cite{Hawking:1981fz} instanton. Denote the position of the (meta)stable false vacuum of the volume moduli described in the preceding section with $t_0\,,\,t_{i,0}$ ($i=2\ldots h^{1,1}$, at vanishing axion VEVs), and its vacuum energy by $V_0$. There is a finite barrier in the direction of the large volume modulus $t={\rm Re}\,T$ at the position $t_B>t_0$, while the smaller blow-up volume directions in K\"ahler moduli space $t_i$ do not show a finite barrier in finite distance from their false vacuum locus $t_{i,0}$. The barrier in the $t$-direction protects against classical decompactification by runaway if $V_0>0$, as there is a de-compactified 10d Minkowski minimum at $t\to\infty$. In our situation, where $t\gg t_i$ at the two extrema of the potential with respect to $t$, we can derive the canonically normalized field $\varphi$ corresponding to $t$ as
\beq
\varphi=\sqrt{\frac32}\,\ln\,t
\eeq
up to small corrections of ${\cal O}(t_i/t)$. As e.g. visible in Figure~\ref{fig_CondOnC}, we can thus approximate the canonically normalized barrier width $\Delta\varphi$ as twice the distance between the false vacuum and the barrier top
\beq
\Delta\varphi\simeq 2\, \sqrt{\frac32}\,(\ln\,t_B\,-\,\ln\,t_0)=\sqrt{6}\,\ln\frac{t_B}{t_0}\quad.
\eeq
The false vacuum is tuned to be a dS vacuum and for purposes of our late-time cosmology should be further tuned to yield an exponentially small positive vacuum energy. Thus the energy difference $\Delta V$ between this false vacuum and the Minkowski vacuum at $t\to\infty$ is exponentially small compared to the barrier height
\beq
\Delta V=V_0\ll V_B\quad.
\eeq
This places us deeply inside the validity regime of the thin-wall approximation to CdL tunneling~\cite{Coleman:1977py,Callan:1977pt,Coleman:1980aw}. In the thin-wall limit tunneling mediated by the CdL instanton gives a decay rate~\cite{Coleman:1980aw}
\beq
\Gamma_{CdL}\sim e^{\frac{S_E(\varphi_0)}{(1+4V_0/3 T^2)^2}}\quad.
\eeq
The term in the denominator of the exponent is called the gravitational suppression factor. Here $S_E(\varphi_0)=-\int d^4x\sqrt{-g}V(\varphi_0)$ denotes the Euclidean action of the pure false vacuum dS space solution at $t_0\,,\,t_{i,0}$ ($i=2\ldots h^{1,1}$) which is~\cite{Kachru:2003aw}
\beq
S_E(\varphi_0)=-\,\frac{24\pi^2}{V_0}<0\quad.
\eeq
$T$ denotes the tension of the CdL bubble of true vacuum which in the thin-wall approximation is given by
\beq
T=\int_{\varphi_0}^\infty d\varphi\sqrt{2\,V(\varphi)}\simeq\sqrt{2\,V_B}\Delta\varphi\simeq6\sqrt{2\,V_B}\,\ln\frac{t_B}{t_0}\quad.
\eeq
We see that the gravitational correction in the decay rate is negligible only if 
\beq
\Delta\varphi \ll \sqrt{\frac{V_0}{V_B}}
\eeq
as for $M_{\rm P}\to\infty$ we have $V_0 M_{\rm P}^2\gg T^2$.
In our case we typically have (see Figure~\ref{fig_CondOnC} again) $\Delta\varphi={\cal O}(0.1)$, so this bound is strongly violated in our own dS vacuum with $V_0\sim 10^{-122}$. So, we are in the opposite situation $V_0 / T^2\ll 1$ and thus to first approximation the CdL decay rate of our class of dS vacua is
\beq\label{CdLdecayrate}
\Gamma_{CdL}\sim e^{-\,\frac{24\pi^2}{V_0}}\cdot e^{\frac{64\pi^2}{T^2}}\sim e^{-10^{122}}\quad.
\eeq
This is extremely long-lived, even compared to our cosmological time scales.

We can now compare this with tunneling mediated by the Hawking-Moss instanton~\cite{Hawking:1981fz} which describes a transition where the field tunnels from the false vacuum first to the top of the barrier, and then classically rolls to the true vacuum. Here, the decay rate comes out to be
\beq
\Gamma_{HM}\sim e^{S_E(\varphi_0)-S_E(\varphi_B)}\sim e^{-\,\frac{24\pi^2}{V_0}+\,\frac{24\pi^2}{V_B}}\quad.
\eeq
The ratio between the two decay rates is~\cite{Kachru:2003aw}
\beq
\frac{\Gamma_{HM}}{\Gamma_{CdL}}\sim e^{24\pi^2\,\left(\frac{1}{V_B}-\frac{4}{T^2}\right)}\quad.
\eeq
Thus, for a sub-Planckian barrier thickness $\Delta\varphi< \sqrt{2} M_{\rm P}$ the HM instanton is sub-dominant to the CdL process, and eq.~\eqref{CdLdecayrate} provides a reasonable estimate for the metastability of our dS vacua.

\section{Stabilization of the K\"ahler modulus and the dilaton} \label{TS_stab}

We now include $S = s + i \sigma$ explicitly into our analysis. Our strategy will be to first determine the supersymmetric locus for $S$ and then include the backreaction from volume stabilization using perturbation theory in the small expansion parameter
\begin{equation}
\frac{\hat\xi}{\hat{\cal V}}\lesssim 0.1\quad
\end{equation}
for typical models. We will use the same logic in section~\ref{TSU_stab_gen} for incorporating the complex structure moduli.

The flux-superpotential has the form $W_0 = C_1 - C_2 \cdot S$, where the $C_1$ and $C_2$ are functions of the complex structure moduli, and the 3-form fluxes. In this section we still assume the complex structure moduli to be integrated out supersymmetrically. The K\"ahler- and superpotential are given as

\begin{eqnarray}
K = K_K + K_{g_s}, \qquad \text{with } && K_K =  - 2 \ln \left( \gamma (T + \bar{T})^{3/2} + \frac{\xi}{2} (S + \bar{S})^{3/2} \right)\,\,,\notag\\
  && K_{g_s} = - \ln \left( S + \bar{S} \right)\,\,,\notag\\
 W = C_1 - C_2 \cdot S + A e^{- a T}\,\,. 
\end{eqnarray}
%
Notice that there is a mixing in the K\"ahler potential due to the $S$ dependence of the parameter $\hat\xi$ controlling the $\alpha'$ correction. The VEV of $s$ has to be chosen large enough, so the string coupling $g_S \equiv 1/ \langle s \rangle$ stays parametrically small. The scalar potential can be organized in the following way:

\begin{eqnarray}
 V(T,S) &=& V^{(T)} + V^{(T,S)} + V^{(S)},\qquad \text{with} \notag\\[3mm]
  V^{(T)} &=& e^{K} \left( K^{T\bar{T}} \left[ W_T \overline{W_T} + (W_T \cdot \overline{W K_T} + c.c) \right] + 3 \hat\xi \frac{\hat{\xi}^2+7\hat{\xi}\hat{\mathcal{V}}+\hat{\mathcal{V}}^2}{(\hat{\mathcal{V}}-\hat{\xi})(\hat{\xi}+2\hat{\mathcal{V}})^2} |W|^2 \right) \,\,,\notag\\
 V^{(T,S)} &=& e^{K} \left( K^{T\bar{S}} D_T W \overline{D_{S} W} + c.c \right) \,\,,\notag\\[3mm]
 V^{(S)} &=& e^{K} \left( K^{S \bar{S}} |D_S W|^2 \right) \,\,.\label{VFTSexact}
\end{eqnarray}
%
The term $V^{(T,S)}$ is due to the mixing of $T$ and $S$ in the K\"ahler potential.

It was shown in \cite{Westphal:2006tn} numerically, that eq.~\eqref{VFTSexact} possesses a meta-stable de Sitter vacuum in the large volume limit eq.~\eqref{approxconditions} with the dilaton being stabilized close to the supersymmetric minimum $D_S W = 0$. We now obtain an analytic understanding of these features using an expansion of eq.~\eqref{VFTSexact} in $\hat{\xi}/\Vol$ and $A e^{- a t}/|W_0|$.


\subsection{Approximating the scalar potential $V(T,S)$ in the large volume limit}

We can calculate $V^{(T)}$ using our results from section $\ref{T_stab}$ and the replacements

\begin{align}
W_0 &\longrightarrow C_1 - C_2 S\,\,,\notag\\
\hat\xi &\longrightarrow \xi (2 s)^{3/2}\,\,,\notag\\
e^{K_K} &\longrightarrow e^{K_K} e^{K_{g_s}} \simeq \left( 2 s \hat{\mathcal{V}}^2 \right)^{-1}\,\,,
\end{align}
%
to obtain the two term potential
\begin{align}
V^{(T)} \simeq& \frac{1}{2 s} \left( \frac{a A e^{-a t} \left[ (C_1 - C_2 s) \cos(a \tau) + C_2 \sigma \sin(a \tau) \right]}{2 \gamma^2 t^2} \right. \notag\\ &+ \left. \frac{3 \xi s^{3/2} \left[(C_1 - C_2 s)^2 + C_2 \sigma^2\right]^2 }{32 \gamma^3 t^{9/2}}\right) \,\,.\label{VT_TS}
\end{align}
To derive $V^{(T,S)}$ and $V^{(S)}$ we have to approximate $K^{T\bar{S}}$ which we find to be 1-st order and $K^{S\bar{S}}$ which is a 0-th order term:
\begin{align}
 V^{(T,S)} \simeq& \frac{ (C_1+C_2 s) \left[ s^{3/2} (-7 C_1+5 C_2 s) \xi +8 A e^{-a t} \gamma t^{3/2} \cos(a \tau) \right]}{64\, s\, \gamma^3 t^{9/2} }\notag\\ 
 &-\frac{7 C_2^2 \sqrt{s}\, \xi\,  \sigma^2}{64\, \gamma^3 t^{9/2}}+\frac{A e^{-a t} C_2\,  \sigma  \sin(a \tau )}{8\, s\, \gamma^2 t^3}\label{VTS_TS}\,\,,\\[4mm]
 V^{(S)} \simeq& \frac{1}{2 s \hat{\mathcal{V}}^2} \left[ (C_1 + C_2 s)^2 + C_2^2 \sigma^2 \right]\,\,.\label{VS_TS}
\end{align}
We see that in this approximate expression for the scalar potential the field $s$ is to 0-th order stabilized by a quadratic potential $(C_1 + C_2 s)^2$ if we neglect terms that are suppressed either by $\xi$ or $e^{-a t}$ relative to the quadratic potential. The supersymmetric locus is

\begin{equation}
 s_0 = - \frac{C_1}{C_2}>0\quad \Rightarrow \quad C_1C_2<0.
\end{equation}

The shift of $s$ to this supersymmetric minimum due to the 1-st order terms $V^{(T)}$ and $V^{(T,S)}$ will be calculated in section $\ref{susybreakingtermsTS}$ to first order. The extremum of $t$ to 1-st order is governed by $V^{(T)}$ only since $V^{(T,S)} \propto D_S W$ and $D_S W$ equals zero to 0-th order so that $V^{(T,S)}$ is actually a 2-nd order term. Finally, the axion field derivatives $V_\tau$ and $V_\sigma$ can be minimized for $\tau = n \pi /a$ for $n \in \mathbb{Z}$ and $\sigma = 0$. As in section $\ref{T_stab}$, we restrict to $\tau = 0$.

\subsection{Moduli Masses}

Using the approximate scalar potential $V(t,s,\tau,\sigma)$ of eq.~\eqref{VFTSexact} we can calculate the mass matrix of the moduli as the second derivative with respect to the real fields. The second derivatives mixing real and imaginary parts vanish exactly at the axionic VEVs $\tau = \sigma = 0$ so the mass matrix is block diagonal.

\begin{equation}
V_{ij} = \left( \begin{array}{cccc}
V_{tt} & V_{ts} & 0 & 0\\
V_{st} & V_{ss} & 0 & 0\\
0 & 0 & V_{\tau\tau} & V_{\tau\sigma}\\
0 & 0 & V_{\sigma\tau} & V_{\sigma\sigma}\\
\end{array} \right)
\label{VijTS}
\end{equation}
Since $V_{tt}$ and $V_{ts}$ are 1-st order and $V_{ss}$ is 0-th order, the eigenvalues of eq.~\eqref{VijTS} are $V_{tt}$ and $V_{ss}$ to 1-st and 0-th order, respectively. 

Next, we note that the kinetic terms of the moduli fields are highly non-canonical. The kinetic part of the Lagrangian reads as
\begin{equation}
{\cal L}=K_{S\bar S}\partial_\mu S\partial^\mu \bar S+K_{T\bar T}\partial_\mu T\partial^\mu \bar T+K_{T\bar S}\left(\partial_\mu T\partial^\mu \bar S+c.c\right)\quad.
\end{equation}
We expand the moduli around their minima in small fluctuations, $S=\langle S\rangle+\delta S$ and $T=\langle T\rangle+\delta T$. Inserting this, we see that in the limit of small fluctuations we get the $K_{I\bar J} (\langle S\rangle,  \langle T\rangle )$ to be constants. In general, one has to diagonalize the K\"ahler metric and then canonically normalize the kinetic terms in the rotated basis of fluctuations, but here we find that in our limit $\hat\xi / \hat{\cal V}\ll 1$ an expansion of the inverse K\"ahler metric $K^{I\bar J}$ shows us that $K^{S\bar T}$ is ${\cal O}(\hat\xi / \hat{\cal V})$ compared to $K^{T\bar T}$ and $K^{S\bar S}$.

Thus, differentiating eq.~\eqref{VFTSexact} and evaluating at $s=s_0$ and $t=t_{min}$, we find for the physical masses

\begin{align}
m_t^2 &\simeq K^{T\bar T}\big|_{\xi=0}V_{tt}=\frac{4 t^2}{3}V_{tt}\notag\\
& \simeq \frac{- 32 a t\, A e^{-a t} C_2 \gamma t^{3/2} (a^2 t^2 + 4 a t +6) + 297  \xi C_1^2 \sqrt{\frac{- C_1}{C_2}} }{48 \gamma^3 t^{9/2}}\bigg|_{t=t_{\text{min}}} \sim \frac{\hat\xi}{\Vol^3}\,\,, \label{mt2approx}\\
%
m_s^2 &\simeq K^{S\bar S}\big|_{\xi=0}V_{ss}=4s^2\,V_{ss} \simeq \frac{- C_1C_2}{2 \gamma^2 t^3}\bigg|_{t=t_{\text{min}}} \sim \frac{1}{\Vol^2}\,\,,\\
m_\tau^2 &\simeq K^{T\bar T}\big|_{\xi=0}V_{\tau\tau} \simeq \frac{(at)^3 A C_2}{2 \gamma^2}\frac{ e^{-a t}}{\gamma^2t^3}\bigg|_{t=t_{\text{min}}}\sim \frac{\hat\xi}{\Vol^3}\,\,,\\ 
m_\sigma^2 &\simeq K^{S\bar S}\big|_{\xi=0}V_{\sigma\sigma} \simeq \frac{- C_1C_2}{2 \gamma^2 t^3}\bigg|_{t=t_{\text{min}}}\sim \frac{1}{\Vol^2}\,\,.
\end{align}
Here we have used that for our dS solutions with $\langle V\rangle \simeq 0$ the product $at\simeq 3$ is roughly constant. In this approximation the fields $s$ and $\sigma$ have the same mass which expresses that they are in the same chiral multiplet and supersymmetry is unbroken in the $S$ direction to 0-th order. Note, that $s$, $\tau$ and $\sigma$ are manifestly positive in our approximation. $t$ could become tachyonic if the exponential term in eq.~\eqref{mt2approx} gets larger than the term proportional to $\xi$. Since $m_t^2 \propto V''(x)$ of eq.~\eqref{VF''x}, a tachyonic direction in $t$ corresponds to a saddle-point of the potential $V(x)$ which is equivalent to violating the upper bound on $C$ eq.~\eqref{dSCondeq} as discussed in section $\ref{dScond}$.

\subsection{Deviation of $s$ from the SUSY minimum and SUSY breaking} \label{susybreakingtermsTS}

In this section, we want to analyze the effect of the 1-st order terms $\delta V^{(1)} \equiv V^{(T)} + V^{(T,S)}$ on the 0-th order potential $V^{(0)} \equiv V^{(S)}$ that stabilizes $s$ in a supersymmetric minimum $s_0 = -C_1/C_2$. We will calculate the shift $\delta s/s_0$ from the supersymmetric minimum $s_0$ due to the 1-st order terms $\delta V^{(1)}$ and show that it is indeed small, i.e. $\order(\hat\xi/\Vol)$. Furthermore, we will show that naturally there appears a hierarchy $m_t^2 \ll m_s^2$ and show that supersymmetry is predominantly broken in the $T$ direction, i.e. $F_T \gg F_S$ where

\begin{equation}
 F_i = e^{K/2} D_i W\,\,.
\end{equation}

Expanding eq.~\eqref{VFTSexact} to first non-vanishing order for zero axionic VEVs $\tau = \sigma = 0$ in $\delta s$ yields

\begin{equation}
V(t,s) = V^{(0)}(t,s_0) + \frac{1}{2} V^{(0)}_{s,s}(t,s_0) (\delta s)^2 + \delta V^{(1)}(t,s_0) + \delta V^{(1)}_s(t,s_0) \delta s + \dots \,\,.
\label{Vtsexpansion_TS}
\end{equation}
Since $s = s_0 + \delta s$ should still be a minimum of the full potential we demand

\begin{align}
\frac{\partial V}{\partial (\delta s)} =0 \quad \Leftrightarrow \quad \delta s &= \frac{- \delta V^{(1)}_s(t,s_0)}{V^{(0)}_{s,s}(t,s_0)}\,\,.\label{deltasTS}
\end{align}
We see from eq.~\eqref{VT_TS} and eq.~\eqref{VTS_TS} that the term $\delta V^{(1)}_s(t,s_0)$ involves terms proportional to $\xi$ and $e^{-a t}$. The latter can be replaced using the condition eq.~\eqref{xmintrans} for the minimum in $t$ at $s=s_0$:

\begin{equation}
 e^{-a t} = \frac{27 \hat\xi\, C_1 }{16\Vol \, A\, a t\, (2+a t)}\,\,.
 \label{expat_replace}
\end{equation}
This yields a function whose $t$ dependence is given by an overall factor $\Vol^{-3}$ and a rational function in $a t = x$. Since we are interested in de Sitter minima with almost vanishing positive cosmological constant we can set $x\simeq 5/2$ according to section $\ref{dScond}$. The mass term $m_s^2$ is obtained solely from $V^{(S)}$ so its $t$-dependence is given by an overall $\Vol^{-2}$ scaling from the overall factor $e^{K}$ in the scalar potential. So finally for the shift we indeed obtain a number of $\order(1)$ times our expansion parameter:\footnote{For a general supergravity analysis of the influence of supersymmetrically stabilized heavy moduli on the stabilization of lighter moduli see \cite{Gallego:2008qi,Gallego:2011jm}, where the ${\cal O}(\xi/\Vol)$ shifts of the heavy moduli were found, too.}

\begin{equation}
\frac{\delta s}{s_0} \simeq \frac{93}{20} \frac{\hat\xi}{\hat{\mathcal{V}}} = 4.65 \frac{\hat\xi}{\hat{\mathcal{V}}}\,\,.
\label{deltas_TS}
\end{equation}

Thus, we have shown that it is consistent to assume the dilaton $s$ to be stabilized approximately supersymmetrically since the 1-st order potential $\delta V^{(1)}$ only has a 1-st order effect on the position of its minimum. Note that the sufficient condition on $C$ for meta-stable de Sitter vacua as it is written down in eq.~\eqref{dSCondeq}, holds for the exact minimum of $s$. If we approximate $s$ by the supersymmetric minimum $s_0$ or to 1-st order by $s_0 + \delta s$ this will slightly change the bounds in eq.~\eqref{dSCondeq}.

We can also use eq.~\eqref{expat_replace} to bring $m_t^2$ into an expression that scales like $\hat\xi / \Vol^{3}$. Setting again $x\simeq 5/2$ we obtain the following hierarchy between the moduli masses

\begin{equation}
 \frac{m_t^2}{m_s^2} \simeq \frac{(a t)^2}{5\,s_0}\cdot\frac{\hat\xi}{\Vol}
\end{equation}
and hence $m_t^2 \ll m_s^2$ parametrically.

Finally, let us calculate the supersymmetry breaking terms $F_T$ and $F_S$. The direction $F_T$ has a non-vanishing 0-th order contribution

\begin{equation}
F_T \simeq - \frac{3 C_1}{\sqrt{-2 C_1 / C_2}\,t \Vol }\,\,.
\end{equation}
As expected, the first non-vanishing contribution to $F_S$ is 1-st order. Other than terms $\propto \hat\xi/\Vol^2$ we have to add a term $\propto (s-s_0)/\Vol$ that we evaluate at $s=s_0 + \delta s$. Inserting eq.~\eqref{deltas_TS} we get

\begin{equation}
 F_S \simeq - \frac{9 C_1 \hat\xi}{10 \sqrt{2}\, \Vol^2 \left(-C_1 / C_2\right)^{3/2}} \simeq -F_T\cdot \frac{3\,t\, C_2 }{10 C_1}\cdot\frac{\hat\xi}{\Vol} 
\end{equation}
so supersymmetry is predominantly broken in the $T$ direction which is what one would expect since $t$ is stabilized in a minimum with spontaneously broken supersymmetry.

The gravitino mass can be approximated to 0-th order to

\begin{equation}
 m_{3/2}^2 = e^{K} |W|^2 \simeq -\,\frac{2\, C_1 C_2}{ \Vol^2}=-\,\frac{\, C_1 C_2}{4\gamma ^2t^3}\sim 10^{-4}\ldots 10^{-3}\quad
\end{equation}
which is of order $\sim M_{\rm GUT}^2$ for typical volumes.

We note that $m_{3/2}< m_s\, , m_\sigma$ which renders the supersymmetric starting point for them a self-consistent approximation. Moreover, the KK scale here is given for a single volume modulus (i.e. no anisotropies are possible) and the volume given in units of $\alpha'$ as $\Vol = L^6$ as
\begin{equation}
m_{KK}=\frac{1}{L\sqrt{\alpha'}}\sim\frac{1}{\Vol^{2/3}}
\end{equation}
while he gravitino mass as well as the moduli masses scale at least $\sim 1/\Vol$. Here we have used the relation between 10d string frame and 4d Einstein frame
\begin{equation}
\frac{1}{\alpha'}=\frac{(2\pi)^7}{2}\,M_{\rm P}^2\,\frac{g_S^2}{\Vol}\quad.
\end{equation}
Therefore, the use of a 4d effective supergravity description is justified, although the separation
\begin{equation}
\frac{m_{3/2}}{m_{KK}}\sim \frac1{\Vol^{1/3}}
\end{equation}
will typically be only of ${\cal O}(0.1)$ here. Nevertheless, there is a parametric hierarchy between the moduli mass scale, the SUSY and the KK-scale in the limit of large volume $\Vol\to\infty$. This suppresses potential mixing between the moduli masses and KK masses alleviating their danger of causing additional tachyonic directions.

We have succeeded now in determining the combined scalar potential of the volume modulus $T$ and the dilaton $S$ in a fully analytical form to first order in a perturbation expansion around the supersymmetric locus for $S$. The resulting full minimum is a tunable dS minimum of the same form and type as found in the previous section for $T$ alone, and it is perturbatively stable under the inclusion of the dynamics of the dilaton $S$.

\subsection{Numerical example}

\begin{table}[t!]
\centering
  \begin{tabular}{c||c|c|c|c|c|c|}
   & $\langle t\rangle$ & $\langle s\rangle$ & $m_t^2$ & $m_s^2$ & $m_\tau^2$ & $m_\sigma^2$\\
  \hline
   ex. & $33.3$ & $7.89$ & $8.2 \cdot 10^{-5}$ & $2.3 \cdot 10^{-3}$ & $2.1 \cdot 10^{-4}$ & $ 1.3\cdot 10^{-3}$\\
   appr. & $32.3$ & $7.92$ & $8.9 \cdot 10^{-5}$ & $ 2.6\cdot 10^{-3}$ & $2.3 \cdot 10^{-4}$ & $ 1.5 \cdot 10^{-3}$\\[0.3cm]
   & $|F_T|$ & $|F_S|$ & $m^{2}_{3/2}$\\
  \cline{1-4}
   ex. & $1.3 \cdot 10^{-3}$ & $3.3 \cdot 10^{-4}$ & $8.3 \cdot 10^{-4}$\\
   appr. & $1.4 \cdot 10^{-3}$ & $2.1 \cdot 10^{-4}$ & $1.2 \cdot 10^{-3}$\\
  \end{tabular}
  \caption{Numerical results for the VEVs in units of $M_P$, mass spectrum and SUSY breaking in units of $M_P^2$, for the parameters of eq.~\eqref{exampleTS}. The exact results are obtained from the full potential for $C_1 = -13.743$, the approximate results are obtained from the approximate potential eq.~\eqref{VT_TS}-\eqref{VS_TS} for $C_1 = -13.926$. In both cases, the field VEVs are calculated by numerical minimization of the respective potential while the moduli masses are the eigenvalues of the second derivative matrix times a factor from canonical field normalization (see text).}
  \label{tab_numTS}
\end{table}

Here, we will shortly compare our previous analytic results for combined $T$, $S$ stabilization to the exact results that one obtains by analyzing the full scalar potential. For concreteness, we will again use the numerical example of~\cite{Westphal:2006tn}. In this example (see eq.~\eqref{exampleT}), we had to fix the constants $W_0$ and $\hat\xi$ which are now given by the flux constants $C_1$ and $C_2$ to 0-th order via

\begin{equation}
 W_0 = 2\,C_1, \qquad \qquad \hat\xi = \xi \left(\frac{-2 C_1}{C_2} \right)^{3/2}\,\,.
\end{equation}
For comparison, we choose a set of parameters from section 4.1 of \cite{Westphal:2006tn}, i.e.

\begin{equation}
 a = \frac{2 \pi}{100},\quad A = 1,\quad \gamma = \frac{\sqrt{3}}{2 \sqrt{5}},\quad \xi = 0.17133,\quad C_1 = -13.743,\quad C_2 = 1.4\,\,.
 \label{exampleTS}
\end{equation}
The choice for $\gamma$ and $\xi$ again corresponds to the quintic $C\mathbb{P}^4_{1,1,1,1,1}$. For this choice of parameters we have $W_0 = -27.49$, $s_0 = 9.9$ and $\hat\xi = 14.9$.

\begin{figure}[t!]
\centering
\vskip -1.5cm
\includegraphics[width= \linewidth]{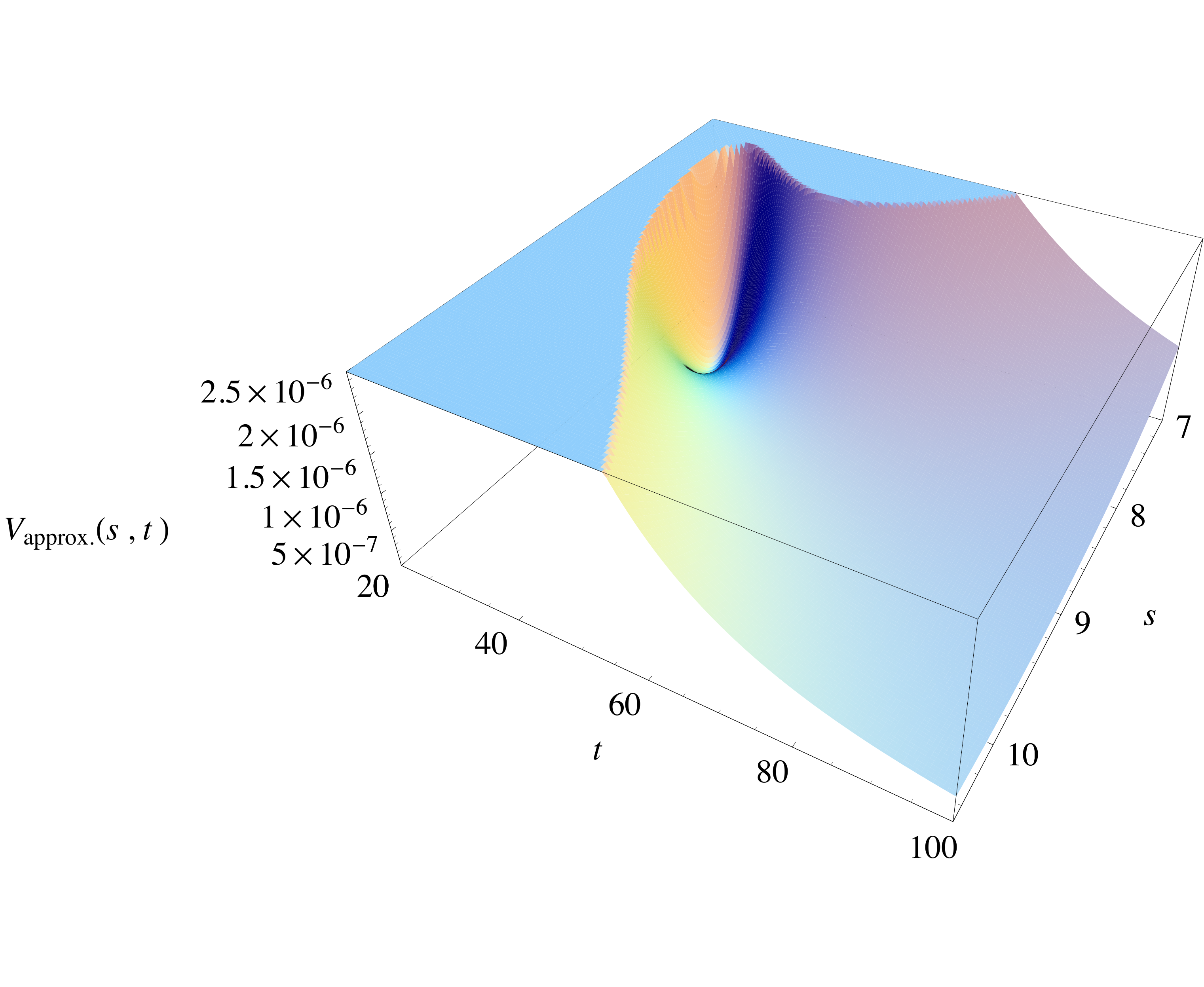}
\vskip -1.5cm
\caption{The approximate potential as a function of $t$ and $s$. Parameter choice were $C_1 = -13.926$, $C_2=1.4$.}
\label{fig_VexactVapproxTS}
\end{figure}

We find the minimum in the $t$ direction of the full potential to lie at $t \simeq 30$. Hence, our expansion parameters are small:

\begin{equation}
 \frac{\hat\xi}{\Vol} \simeq 0.08 \ll 1, \qquad \frac{A e^{-a t}}{|W_0|} \simeq 0.006 \ll 1 \,\,.
\end{equation}
Figure~\ref{fig_VexactVapproxTS} compares the shape of the full potential and the approximate potential, while table~\ref{tab_numTS} presents a summary of the numerical results. The moduli masses show best agreement in the axionic sector. 

This finishes our numerical analysis. Having included the dilaton manifestly, we are now led to the inclusion of the remaining fields missing so far in the full analysis, the complex structure moduli, to which we now turn.



\section{Inclusion of complex structure moduli: general analysis} \label{TSU_stab_gen}

We will now go the final step and include an arbitrary number $h^{2,1}$ of complex structure moduli $U_i = u_i + i \nu_i$ into our stabilization analysis. A commonly used example of a Calabi-Yau 3-fold with one K\"ahler modulus are smooth hypersurfaces in $C\mathbb{P}^4$, for instance the quintic $C\mathbb{P}^4_{1,1,1,1,1}$. In this case, we generically have $\mathcal{O}(100)$ complex structure moduli so the Euler number of our Calabi-Yau 3-fold will be of the order

\begin{equation}
 \chi = 2 (h^{1,1}-h^{2,1}) \sim \mathcal{O}(-200)\,\,.
\end{equation}

The analysis of the previous sections led us to expect the leading $\alpha'$ correction to the K\"ahler potential to be $\hat\xi = \mathcal{O}(10)$. This needs the dilaton $\text{Re}(S) = g_S^{-1}$ to be at weak coupling:

\begin{equation}
 \hat\xi = - \frac{\zeta(3)}{4 \,\sqrt{2}\, (2 \pi)^3} \, \chi \, (2 \, s)^{3/2} \simeq 0.5 \, g_S^{-3/2} \quad \Rightarrow \quad g_S \simeq \mathcal{O}(0.1)\,\,.
\end{equation}

\noindent Finding meta-stable minima of an effective scalar potential of $\mathcal{O}(100)$ complex scalar fields is in general a challenging and cumbersome task. A further difficulty enters by the fact that the explicit form of the K\"ahler potential and the superpotential

\begin{align}
K &= K_K + K_{g_s} + K_{c.s.}, \quad \text{with } &&K_K = - 2 \ln \left( \gamma (T + \bar{T})^{3/2} + \frac{\hat\xi(S,\bar{S})}{2} \right)\,\,,\notag\\
 & &&K_{g_s} = - \ln \left( S + \bar{S} \right)\,\,,\notag\\
 & &&K_{c.s.} = - \ln \left( - i \int_{CY_3} \bar{\Omega}(\bar{U}_i) \wedge \Omega(U_i) \right)\,\,,\label{KTSU}\\
W &= C_1(U_i) - C_2(U_i) \cdot S + A e^{- a T}\,\,,&\label{WTSU}
\end{align}

\noindent of the complex structure sector are only known explicitly for some special Calabi-Yau threefolds~\cite{Candelas:1990rm}. We neglect the dependence of $A$ on the complex structure sector and assume it to be constant. Note that $A$ always comes together with an exponential term $e^{-a T}$ in the superpotential and hence also in the scalar potential. Thus, the effect of a non-trivial dependence of $A=A(z_i)$ on the complex structure moduli stabilization will effectively be suppressed by an overall factor $\hat\xi / \Vol$. However, since in general the function $A(z_i)$ is not known, we cannot go beyond this qualitative argument in a model-independent way. This leaves us with a possible caveat, as a very steep  functional dependence of $A(z_i)$ might derail our perturbative treatment of complex structure moduli stabilization in certain examples.

Similar to eq.~\eqref{VFTSexact}, we can split the full scalar potential into four parts

\begin{equation}
 V = V^{(T)} + V^{(T,S)} + V^{(S)} + V^{(U)} 
 \label{V4sum}
\end{equation}
where $V^{(T)}$ contains the F-terms of $T$ and the $-3 |W|^2$ term and $V^{(S)}$ and $V^{(U)}$ are the F-terms of $S$ and the $U_i$, respectively and $V^{(T,S)}$ mixes the F-terms of $T$ and $S$. The first three terms of~\eqref{V4sum} are given in~\eqref{VFTSexact} while $V^{(U)}$ is given by

\begin{equation}
 V^{(U)} = e^K K^{U_i \bar{U}_j} D_{U_i} W \overline{D_{U_j} W}\,\,.
 \label{VUdetailed}
\end{equation}

From our analysis in section $\ref{TS_stab}$, we expect a meta-stable minimum of the effective scalar potential which includes the complex structure moduli to have the following properties: The complex structure moduli should be stabilized approximately in a supersymmetric minimum like the dilaton since they enter the scalar potential similarly. They are even further decoupled from the SUSY breaking K\"ahler modulus since there is no mixing term in the K\"ahler potential for the complex structure moduli. We will show in section $\ref{susybreaking_TSUgen}$ that the deviation is in general a 1-st order effect and hence the fields are stabilized supersymmetrically to 0-th order.

\subsection{Deviation of $s$ and $u_i$ from the SUSY minimum} \label{susybreaking_TSUgen}

In this section, we repeat the analysis of section $\ref{susybreakingtermsTS}$ for the additional inclusion of the complex structure moduli. The 1-st order terms of the scalar potential include terms that are proportional to either $e^{-a t}$ or $\hat\xi$ so we write it as a perturbance $\delta V^{(1)} = V^{(T)} + V^{(T,S)}$ of the 0-th order scalar potential $V^{(0)} = V^{(S)} + V^{(U)}$. Expanding this to first non-vanishing order in $\vec{\theta}=(s,u_i)$ around the supersymmetric minimum $\vec{\theta}_0=(s_0,u_{0i})$ gives

\begin{equation}\label{pertexpansion}
 V = V^{(0)} + \frac{1}{2} \underbrace{(\vec{\theta}-\vec{\theta}_0)}_{\delta \vec{\theta}} V^{(0)}_{\vec{\theta}_0 \,\vec{\theta}_0} (\vec{\theta}-\vec{\theta}_0) + \delta V^{(1)} + \delta V^{(1)}_{\vec{\theta}_0} (\vec{\theta}-\vec{\theta}_0) + \dots \,\,,
\end{equation}

\noindent where subscript $\vec{\theta}_0$ denotes differentiating with respect to $\vec{\theta}$, evaluated at $\vec{\theta}_0$. Notice, that we again only expand around the real parts of the moduli fields since the supersymmetric minimum for all axionic VEVs equal to zero is an exact minimum of the scalar potential. Demanding $\delta \vec{\theta}$ to still be a minimum of $V$ we get an expression for $\delta \vec{\theta}$ in terms of 0-th order terms that is similar to eq.~\eqref{deltasTS}:

\begin{equation}
V_{\delta \vec{\theta}} =0 \quad \Leftrightarrow \quad \delta \vec{\theta} = - \left( V^{(0)}_{\vec{\theta}_0 \,\vec{\theta}_0}\right)^{-1} \cdot \delta V^{(1)}_{\vec{\theta}_0}\,\,. \label{deltatheta}
\end{equation}

We will now estimate the correction $\delta \vec{\theta}$ for a general complex structure sector to be of the order $\hat\xi/\Vol$ multiplied with terms depending on $K_{c.s.}$, $W_0$ and derivatives of these expressions with respect to $s$ and $u_i$. First, let us note that the matrix $V^{(0)}_{\vec{\theta}_0 \,\vec{\theta}_0}$ has to be positive definite. It is not sufficient to demand the weaker condition of Breitenlohner-Freedman vacuum stability~\cite{Breitenlohner:1982bm} since we are spontaneously breaking supersymmetry in the $T$ direction to obtain a de Sitter vacuum. Hence the feature of AdS space that keeps a tachyon from exponentially rolling down a negative definite $V^{(0)}_{\vec{\theta}_0 \,\vec{\theta}_0}$ is absent in our case. To analyze the scaling of $V^{(0)}_{\vec{\theta}_0 \,\vec{\theta}_0}$ with respect to our expansion parameter $\hat\xi/\Vol$ only the overall factor $e^{K}$ is relevant since there is otherwise no $t$ dependence in $V^{(0)}$. Hence 
\begin{equation}
 V^{(0)}_{\vec{\theta}_0 \,\vec{\theta}_0} \sim \Vol^{-2}\quad.
\end{equation}

To analyze the scaling of $\delta V^{(1)}_{\vec{\theta}_0}$ with respect to $\hat\xi/\Vol$, we have to build the derivatives of $V^{(T)}$ and $V^{(T,S)}$ with respect to $s$ and $u_i$ respectively and evaluate at the supersymmetric minimum. Note that it is not a priori clear that since $V^{(T,S)}$ scales with $\hat\xi/\Vol^3$ this also applies to the derivative of $V_{TS}$ with respect to $\vec{\theta}$. For the derivatives of $V^{(T)}$, we can replace the term proportional to $e^{-a t}$ by an expression in $\hat\xi/\Vol$ using the $t$ minimum condition eq.~\eqref{xmintrans} after differentiation. Furthermore, we use $V^{(T)} \simeq 0$ at the minimum of $t$, i.e. de Sitter, and $a t \simeq 5/2$ to obtain

\begin{align}
 V^{(T)}_{s} &= \frac{3}{16\,s^2} e^{K_{c.s.}} (3 W_0 + 2 s (W_0)_{s}) W_0 \frac{\hat\xi}{\Vol^3}\,\,,\notag\\
 V^{(T)}_{u_i} &= \frac{3}{8\,s} e^{K_{c.s.}} (W_0)_{u_i} W_0 \frac{\hat\xi}{\Vol^3}\,\,.
\end{align}
To calculate the derivatives of $V^{(T,S)}$, note that $V^{(T,S)}$ in eq.~\eqref{VTS_TS} can be brought into the form

\begin{equation}
 V^{(T,S)} = - e^{K_{c.s.}} D_S W_0 \left( \frac{(D_S W_0 \,s- 3 W_0)\,\hat\xi}{2\Vol^3} + \frac{2 A e^{-a t}}{\Vol^2} \right) \sim \frac{\hat\xi^2}{\Vol^4}\,\,,
 \label{VTSDSW0}
\end{equation}
by using the identities for the 0-th order covariant derivative and the superpotential

\begin{align}
D_S W_0 &= -\frac{C_1+C_2 s}{2\,s}\,\,,\notag\\  
W_0 &= C_1 - C_2 s \,\,,
\end{align}
to replace the parameters $C_1$ and $C_2$ in eq.~\eqref{VTS_TS}. Differentiating with respect to $\vec{\theta}$ and afterwards setting $D_S W_0 = 0$ and replacing terms proportional to $e^{-a t}$ using eq.~\eqref{xmintrans}, only the following term survives:

\begin{equation}
 V^{(T,S)}_{\vec{\theta}_0} = \frac{33}{20}\, e^{K_{c.s.}} W_0 (D_S W_0)_{\vec{\theta}_0} \frac{\hat\xi}{\Vol^3}\,\,.
\end{equation}
After calculating the derivatives $(D_S W_0)_{s}$ and $(D_S W_0)_{u_i}$ we finally obtain

\begin{align}
 \delta V^{(1)}_{s_0} &= \frac{3\,W_0}{80\,s^2}\, e^{K_{c.s.}} \left( 37 W_0 - 12 s\, (W_0)_s \right) \frac{\hat\xi}{\Vol^3}\,\,,\notag\\
 \delta V^{(1)}_{u_{i0}} &= \frac{3\,W_0}{20\,s}\, e^{K_{c.s.}} \left( 11 s\, (W_0)_{S\,u_i} - 3\, (W_0)_{u_i} \right) \frac{\hat\xi}{\Vol^3}\,\,.
\end{align}

We conclude that $\delta V^{(1)}_{\vec{\theta}_0}$ scales as a product of $W_0$ and an expression of derivatives of $W_0$. Both terms scale linearly in the flux quanta of $G_{(3)}$. We also expect $V^{(0)}_{\vec{\theta}_0 \,\vec{\theta}_0}$ to scale quadratically in the flux quanta of $G_{(3)}$, due to differentiating twice with respect to $\vec{\theta}$. So finally going back to eq.~\eqref{deltatheta} we indeed obtain

\begin{equation}
 \delta \vec{\theta}_i \sim \frac{\hat\xi}{\Vol}
 \label{vecthetascaling}
\end{equation}
to be a 1-st order perturbation of the supersymmetric minimum $\vec{\theta}_0$. The scaling of $\delta \vec{\theta}$ described in eq.~\eqref{vecthetascaling} induces the scaling of the covariant derivatives in $\vec{\theta}$

\begin{equation}
 D_i W \simeq (D_i W_0)_{\vec{\theta}_0} \cdot \delta \vec{\theta} \sim \frac{\hat\xi}{\Vol} \qquad \text{for } i=s,u_1,\dots,u_{h^{2,1}}\,\,,
\end{equation}
since $D_i W =0$ at 0-th order and all components of $\delta \vec{\theta}$ scale with $\hat\xi/\Vol$. For the $F_i$ terms, this implies a scaling $\propto \hat\xi/\Vol^2$.

Note that our analysis does not take into account a possible dependence of $\delta \vec{\theta}$ on $h^{2,1}$. This implies the potential caveat that a perturbative expansion of the shift from the supersymmetric minimum in $\hat\xi/\Vol$ might not be consistent for large $h^{2,1}$, as we will now discuss. The parts of the scalar potential $V^{(T)}$ and $V^{(T,S)}$ depend on $U_1,\dots,U_{h^{2,1}}$ via the flux superpotential $W_0$. Hence, when we calculate the deviation of the 0-th order supersymmetric VEV of the dilaton or a complex structure modulus along the lines of section $\ref{susybreakingtermsTS}$ to 1-st order we might expect the deviation to depend on the number of fields that are supersymmetrically stabilized. In the worst case, one could expect the deviation to grow with the number of fields included such that the 1-st order deviation would eventually become of the same order as the 0-th order VEV which would make our perturbative expansion valid only up to certain number of fields included. This is what one could expect naively, since a growing number of fields could 'pull away' the supersymmetrically stabilized fields from their VEVs via $\delta V^{(1)}$ the stronger the more fields are included.

However, we will give here a short argument why we expect no such deleterious dependence of the shifts $\delta \vec{\theta}$ on $h^{2,1}$ to arise. Upon inspection of eq.~\ref{deltatheta} concerning the $u_i={\rm Re}\,U_i$ we see that we can approximate the mass matrix $V_{\vec\theta_0 \vec\theta_0}^{(0)}$ entering there by two extreme cases within which we will typically find realistic examples.

Consider first the non-generic case, where $V_{\vec\theta_0 \vec\theta_0}^{(0)}\sim \langle\mu^2\rangle {\rm diag}({\cal O}(1),\ldots,{\cal O}(1))$ is roughly diagonal, where $\mu$ denotes the common mass scale assumed for this non-generic case. Now we note that $\delta V^{(1)}_{\vec\theta_0}\sim |(W_0)_{\vec\theta_0}|$ and from the 3-cycle decomposition of the CY threefold we have
\begin{eqnarray}\label{CSsuperpot}
W_0=\frac{1}{2\pi}\int_{CY_3}G_{(3)}\wedge \Omega &\sim & \sum_{i=1}^{h^{2,1}}\left(\int_{A_i}G_{(3)}\int_{B^i}\Omega+\int_{B^i}G_{(3)}\int_{A_i}\Omega\right)\nonumber\\
&=&\;\;\quad\sum_{i=1}^{h^{2,1}}\left( N_i\, \Pi^i(U_j)\;\;\quad+\quad M^i\, U_i\right)\;.
\end{eqnarray}
Here the $\Pi^i(U_j)$ denote the periods of the CY, the complex structure coordinates Poincare dual to the $U_a$. At a generic point in the interior of moduli space of a generic CY we expect the periods, in being the dual complex structures, to have the same sizes as the $U_i$, and thus $\delta V^{(1)}_{\vec\theta_0}\sim |(W_0)_{\vec\theta_0}|$ will be roughly constant in $h^{2,1}$. For our first case of a roughly diagonal mass matrix this implies that the shifts $\delta\vec\theta$ are roughly constant in $h^{2,1}$.

Now consider the 2nd generic case of a non-diagonal mass matrix which we approximate by $V_{\vec\theta_0 \vec\theta_0}^{(0)}~\sim \langle\mu^2\rangle{\cal O}(1)$ $\forall i,j=1\ldots h^{2,1}$. In this case, each row on the LHS of eq.~\eqref{deltathetabeforeinvert}, which is eq.~\eqref{deltatheta} before inversion, contains a sum over all $\delta \vec{\theta}_i$ with roughly equally sized coefficients. 
\begin{equation}\label{deltathetabeforeinvert}
 V^{(0)}_{\vec{\theta}_0 \,\vec{\theta}_0} \cdot \delta \vec{\theta} = - \delta V^{(1)}_{\vec{\theta}_0}
\end{equation}
Now as $V_{\vec\theta_0 \vec\theta_0}^{(0)}$ has roughly equal sized entries everywhere, eq.~\eqref{deltathetabeforeinvert} should have a solution
 \begin{equation}
 \delta\vec\theta\sim \left\langle\frac{1}{\mu^2}\right\rangle\,\frac{1}{h^{2,1}}
 \end{equation}
 for the shifts of the complex structure moduli,  where $\mu$ denotes the mass scale of the eigenvalues of a mass matrix with roughly equal entries everywhere. As a given tree-level mass matrix  $V_{\vec\theta_0 \vec\theta_0}^{(0)}$ for a given model will in general fall in between these two extreme cases, we expect  no positive power of $h^{2,1}$ to appear in the shifts $\delta\vec\theta$.

We will supplement this line of thinking by an explicit example based on $T^6$. This is presented in section~\ref{TSU_stab}, and we will show there that, in fact, the dependence is harmless as there we will have $\delta s\sim const.$, and $\delta\vec u\sim 1/h^{2,1}$.

We finally note in passing, that the structure of the complex structure superpotential, eq.~\eqref{CSsuperpot}, and the corresponding K\"ahler potential eq.~\eqref{KTSU}
\beq
K_{c.s.}=-\ln\left(U_i\bar\Pi^i(U_j)-\Pi^i(U_j)\bar U_i\right)
\eeq
ensure, similarly to the case for $S$ in its potential eq.~\eqref{VS_TS}, that $e^K$ does not contain more inverse powers of the $U_i$ than the F-terms $K^{i\bar\jmath}D_{U_i}W\overline{D_{U_j}W}$. This implies, that there is no finite potential barrier in finite field space distance in the space of the complex structure moduli and the dilaton which separates the flux vacuum locus from a possible  Minkowski minimum at large ${\rm Re}\,U_i$ or ${\rm Re}\,S$. Instead, the scalar potential of the complex structure moduli and the axio-dilaton grows at large distance without limit. This justifies their neglect in the treatment of vacuum decay in Section~\ref{decay}.

\subsection{Backreaction on the K\"ahler modulus} \label{kahlerback_TSUgen}

We will now derive an expression for the 1-st order shift in $\delta t$ of the K\"ahler modulus due to 2-nd order terms in the scalar potential. $\delta t$ will then be used to calculate the perturbance of the mass $m_t^2 \simeq K^{T\bar{T}}\cdot V_{tt}^{(T)}$ due to these 2-nd order terms.

Splitting eq.~\eqref{V4sum} into 1-st order $V^{(T)}$ and 2-nd order $\delta V^{(2)} = V^{(T,S)} + V^{(S)} + V^{(U)}$ terms we can perform an expansion in $\delta t$ along the lines of eq.~\eqref{Vtsexpansion_TS}-\eqref{deltasTS} in $\delta t$ and obtain

\begin{equation}
 \delta t = - \frac{(\delta V^{(2)})_t}{V_{tt}^{(T)}}\,\,.
 \label{del2Vmt_gen}
\end{equation}
The scaling of $V^{(S)}$ and $V^{(U)}$ is $(\hat\xi/\Vol)^2$ from the $|D_i W|^2$ term, times an $1/\Vol^2$ from the overall factor $e^{K}$. Evaluating $V^{(T,S)}$ to 2-nd order we can make use of eq.~\eqref{VTSDSW0} in only keeping terms linear in $D_S W_0$. So we get

\begin{equation}
 \delta V^{(2)} \sim \frac{\hat\xi^2}{\Vol^4}\,\,.
\end{equation}
which additionally depends quadratically on the flux quanta. Effectively, all $t$ dependence of $\delta V^{(2)}$ is captured in an overall factor $1/\Vol^4$ so that differentiating with respect to $t$ will just give an overall factor $\propto -1/t$. The expression for $V_{tt}^{(T)}$ was calculated in eq.~\eqref{mt2approx}. It scales quadratically in the flux quanta since it is proportional to $W_0^2$. Inserting into eq.~\eqref{del2Vmt_gen}, we obtain

\begin{equation}
 \frac{\delta t}{t} = \frac{ \hat\xi \Delta}{\Vol}\,\,,
\end{equation}
where $\Delta$ is a function which is $\order(1)$ in the fluxes, whose overall sign and dependence on $h^{2,1}$ and hence the smallness of $\delta t/t$ is in general unknown.

We can expand the perturbed mass $\tilde{m}_t^2$

\begin{align}
 \tilde{m}_t^2 &= m^2_t + (\partial_t m^2_t) \, \delta t + \frac{1}{2} (\partial_t^2 m^2_t) \, \delta t^2 + \dots\notag \\
 &=  \frac{5 W_0^2 }{4 s \, \Vol^2}\cdot\frac{\hat\xi}{\Vol}\; e^{K_{c.s.}}\,\left(1 - \frac{31}{2} \frac{\hat\xi \Delta}{\Vol} + \order\left( \frac{\hat\xi \Delta}{\Vol}\right)^2 \right)\,\,.
\end{align}
So if $\Delta$ is negative it cannot cause a tachyonic direction in $t$. However, if $\Delta$ is positive, only values of $\Delta$ that are smaller than roughly $\order(10)$ can be allowed to keep the spectrum tachyon free. Due to its constant scaling in the fluxes we typically expect $\Delta = \order(1)$.

Let us pause here again to discuss a possible dependence of the expansion on $h^{2,1}$. Once it is shown that the dilaton and complex structure moduli are stabilized supersymmetrically with a 1-st order deviation one expects this to induce a 2-nd order term in the potential. This is due to the quadratic dependence of $V^{(S)}$ and $V^{(U)}$ on the respective F-terms and the structure of $V^{(T,S)}$ which is a 1-st order term times $F_S$. Since $V^{(T)}$ is a 1-st order term we expect an effective 1-st order correction on the stabilization of $t$. A correction of the VEV of $t$ induces a correction in $m_t^2$ which could in the worst-case create a tachyonic direction in $t$.

Similar to the situation discussed above for the deviation of the dilaton and the complex structure moduli from the supersymmetric minimum, there is the danger that the correction to $m_t^2$ will be negative and scale with positive powers of $h^{2,1}$. Then, a non-tachyonic $t$ direction would only be possible up to a certain upper bound on $h^{2,1}$. Note, that in case the correction to $m_t^2$ is positive, a scaling with $h^{2,1}$ would even increase $m_t^2$ and make this direction more stable in the end. 

At this point, we have succeeded now in determining the combined scalar potential of the volume modulus $T$,the dilaton $S$, and an arbitrary number $h^{2,1}$ of complex structure moduli $U_i$ in a fully analytical form to first order in a perturbation expansion around the supersymmetric locus for the $S, U_i$. The resulting full minimum is a tunable dS minimum of the same form and type as found in the previous section for $T$ or $T$ and $S$, and it is perturbatively stable under the inclusion of the dynamics of the dilaton $S$ and all $U_i$ (with certain caveats, as there may be non-generic dependence on $h^{2,1}$ in the coefficients of the perturbation expansion).

\section{Inclusion of complex structure moduli: concrete toy example} \label{TSU_stab}

We will now work out the dependence of the 1-st order deviation from the supersymmetric minimum and the 2-nd order K\"ahler modulus backreaction on $h^{2,1}$ for a concrete choice of the K\"ahler potential and superpotential for the complex structure sector. Our guiding example will be the complex structure of a (possible orbifolded) $T^6$ orientifold compactification.

We will show that here the 1-st order shifts from the supersymmetric minimum are actually either independent of $h^{2,1}$ or even decrease with negative powers of $h^{2,1}$ for our specific choice of $K_{c.s.}$ and $W_0$. Furthermore, we will show that the backreaction on the K\"ahler modulus will not introduce a tachyon. This means that our construction: a K\"ahler modulus stabilized by the interplay of the leading $\alpha'$ correction and non-perturbative effects together with approximately supersymmetrically flux stabilized dilaton and complex structure moduli can contain meta-stable de Sitter vacua for an arbitrary large value of $h^{2,1}$ in this toy model. We will show this by explicitly calculating the minima of the scalar potential for the K\"ahler modulus, the dilaton and $h^{2,1}$ complex structure moduli.

Our guidance from the example of $T^6$ gives us an Ansatz for complex structure sector 

\begin{align}
 K_{c.s.} &= - \ln \left( - i \int_{CY_3} \bar{\Omega} \wedge \Omega \right) = - \sum_{i=1}^{h^{2,1}} \ln\left( U_i + \bar{U}_i\right)\,\,,\label{cs_KAnsatz}\\
 W_0 &= c_1 + \sum_{i=1}^{h^{2,1}} d_{1i} U_i - (c_2 + \sum_{i=1}^{h^{2,1}} d_{2i} U_i) \cdot S\,\,,
\label{cs_WAnsatz}
\end{align}

\noindent with the flux constants $c_i, \,d_{ij} \in \mathbb{R}$. The structure above has been found for the various orientifolded orbifolds of $T^6$ discussed in \cite{Lust:2005dy}. The toroidal orbifold-orientifold are orbifold limits of non-trivial CY threefolds, yet at the orbifold point they preserve the simple structure of K\"ahler potential of the untwisted complex structure moduli inherited from $T^6$, which enables us to do explicit calculations. Explicating the arguments of the previous chapter on a general CY threefold compactification requires knowledge of the periods of the threefold, which in general is not available.


\subsection{The supersymmetric minimum for the dilaton and the complex structure moduli}

We now want to calculate the position of the supersymmetric VEVs of the dilaton and complex structure moduli which corresponds to their 0-th order VEV when the K\"ahler modulus is included in the stabilization. We have the $D_i W = W_i + K_i W$ that follow from eq.s~\eqref{cs_KAnsatz}, \eqref{cs_WAnsatz}:

\begin{align}
 D_S W &= - \frac{c_1 + \sum_i d_{1i} U_i + (c_2 + \sum_i d_{2i} U_i)\,\bar{S}}{S + \bar{S}}\,\,,\notag\\
 D_{U_i} W &= - \frac{c_1 + \sum_{j\neq i} d_{1j} U_j - d_{1i} \bar{U}_i + (c_2 + \sum_{j\neq i} d_{2j} U_j - d_{2i} \bar{U}_i)\,S}{U_i + \bar{U}_i}\,\,.\label{DUiW}
\end{align}
To obtain the supersymmetric minima we need to solve

\begin{align}
 \text{Re}(D_S W) &= \text{Re}(D_{U_i} W) = 0\,\,,\label{DSUiW}\\ 
 \text{Im}(D_S W) &= \text{Im}(D_{U_i} W) = 0\,\,.\label{barDSUiW}
\end{align}
We see that due to $c_i \in \mathbb{R}$ setting $\nu_i \equiv \text{Im}(U_i)$ and $\sigma \equiv \text{Im}(S)$ to zero will always be a solution of the eq.s~\eqref{barDSUiW}.

\subsubsection{Solving for real parts} \label{repartsSUSYsol}

We now have to solve the equations

\begin{align}
 - \frac{c_1 + \sum_i d_{1i} u_i + (c_2 + \sum_i d_{2i} u_i)\,s}{2\,s} &= 0\,\,,\notag\\
 - \frac{c_1 + \sum_{j\neq i} d_{1j} u_j - d_{1i} u_i - (c_2 + \sum_{j\neq i} d_{2j} u_j - d_{2i} u_i)\,s}{2\,u_i} &= 0 \,\,.
\end{align}
In general, an analytic solution of these $h^{2,1}+1$ non-linear equations is difficult to obtain. However, if we restrict the flux parameters to $d_{1i} = d_{2i} \equiv d_i$ the $h^{2,1}$ equations $\text{Re}(D_{U_i} W) = 0$ obtain a symmetric structure in $d_i u_i$. The solution will always respect the condition

\begin{equation}
 d_1 u_1 = d_2 u_2 = \dots = d_{h^{2,1}} u_{h^{2,1}} \equiv d u \,\,.\label{duequal}
\end{equation}
Hence, the $h^{2,1}$ equations $\text{Re}(D_{U_i} W) = 0$ are all equivalent and effectively only two equations remain:

\begin{align}
 - \frac{c_1 + h^{2,1} \,d u + (c_2 + h^{2,1} \,d u)\,s}{2\,s} &= 0\,\,,\notag\\
 - \frac{c_1 + (h^{2,1}-2) \,d u - (c_2 + (h^{2,1}-2) \,d u)\,s}{2\,u_i}&= 0\,\,.
\end{align}
Now, finding solutions for $s$ and $du$ as functions of $c_1, c_2$ and $h^{2,1}$ reduces to solving a quadratic equation. The detailed form of these expressions is not instructive for our analysis. If instead we solve for the flux constants $c_1$ and $c_2$ as a function of $s, \,du$ and $h^{2,1}$

\begin{align}
 c_1 &=  - (h^{2,1} -1 + s)\,du\,\,,\label{c1dus}\\
 c_2 &= -\frac{1 - s + h^{2,1} s}{s} \,du\,\,,\label{c2dus}
\end{align}

\noindent and insert this result into eq.~\eqref{cs_WAnsatz} we find 

\begin{equation}
W_0 = 2\,du\,(1-s)\,\,.
\label{W0dus}
\end{equation}

This now tells us how to explicitly construct supersymmetric minima for the fields $s$ and $u_i$ that fulfill our sufficient condition eq.~\eqref{dSCondeq} for de Sitter vacua: We choose $W_0$ and $\hat\xi$ (and hence $s$) so that eq.~\eqref{dSCondeq} is fulfilled. Then eq.~\eqref{W0dus} fixes the value of $du$ and eq.s~\eqref{c1dus} and eq.~\eqref{c2dus} determine $c_1$ and $c_2$. For every complex structure modulus, only the product $d_i u_i$ is fixed and the VEVs of the $u_i$ can be chosen by adjusting the parameters $d_i$. Note that for $s=1$ the superpotential eq.~\eqref{W0dus} vanishes. We are not interested in this peculiar VEV of the dilaton since we demand small string coupling, i.e. $s \simeq \order(10)$. So from now on we will always assume $s>1$.

\subsubsection{Solving for imaginary parts} \label{impartsSUSYsol}

We now solve the eq.s~\eqref{barDSUiW} for a more general choice of $\nu_i \neq 0$. However, we still restrict to $\sigma = 0$ so the real and imaginary parts of the complex structure moduli fields decouple:

\begin{align}
 &\text{Im}(D_S W) = - \frac{\sum_i d_{1i} \nu_i + \left(\sum_i d_{2i} \nu_i \right) \, s}{2 s} = 0\,\,,\notag\\
 &\text{Im}(D_{U_i} W) = - \frac{\sum_i d_{1i} \nu_i - \left(\sum_i d_{2i} \nu_i \right) \, s}{2 u_i} = 0\,\,.
\end{align}
In this case the $h^{2,1}$ equations $\text{Im}(D_{U_i} W) = 0$ are manifestly the same other than in the case $\text{Re}(D_{U_i} W) = 0$ where those terms were just highly symmetric. Clearly the two equations are solved by

\begin{equation}
 \sum_i d_{1i} \nu_i = \sum_i d_{2i} \nu_i = 0\,\,.
\end{equation}
These two equations leave $h^{2,1}-2$ of the $\nu_i$ undetermined and in the special case $d_{1i} = d_{2i}$ that was considered in the previous subsection this degeneracy is even increased to $h^{2,1}-1$. This corresponds to $h^{2,1}-1$ flat axionic directions in the scalar potential so we should observe exactly this number of massless axions in our following analysis.

\subsection{Approximating the scalar potential $V(T,S,U_i)$ in the large volume limit}

In the following, we restrict $K_{c.s.}$ and $W_0$ to be of the form that we specified in eq.~\eqref{cs_KAnsatz} and eq.~\eqref{cs_WAnsatz}. Then, the parts $V^{(T)}$, $V^{(T,S)}$ and $V^{(S)}$ are obtained from eq.~\eqref{VT_TS}-\eqref{VS_TS} by the replacements

\begin{align}
C_1 &\longrightarrow c_1 + \sum_i d_{1i} U_i\,\,,\notag\\
C_2 &\longrightarrow c_2 + \sum_i d_{2i} U_i\,\,,\notag\\
e^{K_K+K_{g_s}} &\longrightarrow e^{K_K+K_{g_s}+K_{c.s.}}\,\,.
\end{align}
With these replacements calculating $V^{(T)}$, $V^{(T,S)}$ and $V^{(S)}$ in dependence of the real field components of $T$, $S$ and $U_i$ is straightforward. The only complication arises from the fact that $C_1$ and $C_2$ are complex now in contrast to section $\ref{TS_stab}$ where they were assumed to be real.

The K\"ahler metric of the complex structure sector is diagonal and hence is the inverse

\begin{equation}
 K^{U_i \bar{U}_j} = \text{diag} \left( (U_i + \bar{U}_i)^2 \right)\,\,.
\end{equation}
With $D_{U_i} W$ already calculated in eq.~\eqref{DUiW} we can write down the scalar potential for the complex structure sector:

\begin{align}
 V^{(U)} = &e^K \sum_i \left( \left[ c_1 + \sum_{j \neq i} d_{1j} u_j - d_{1i} u_i - (c_2 + \sum_{j \neq i} d_{2j} u_j - d_{2i} u_i) \,s + ( \sum_j d_{2j} \nu_j )\, \sigma \right]^2 \right.\notag \\
 &+ \left. \left[  \sum_j d_{1j} \nu_j  - ( \sum_j d_{2j} \nu_j ) \,s - (c_2 + \sum_{j \neq i} d_{2j} u_j - d_{2i} u_i) \, \sigma \right]^2 \right) \,\,.\label{VU}
\end{align}

\subsection{Moduli masses}

We now want to find extrema of the scalar potential $V$ and check if the second derivative $V_{ij}$ is a positive definite matrix. This is the case if all eigenvalues of $V_{ij}$, i.e. the moduli masses are positive. We will calculate analytic expressions for the moduli masses, show that they are always positive for the real parts and never negative for the imaginary parts (axions) of the moduli fields in this section. Thus we can exclude tachyonic directions in the scalar potential.

At first, we find that setting all imaginary parts of the moduli to zero

\begin{equation}
 \tau = \sigma = \nu_i = 0
\end{equation}

\noindent is a solution of

\begin{equation}
 V_{\tau} = V_{\sigma} = V_{\nu_i} = 0\quad.
\end{equation}

\noindent This is the same axionic behavior that we found in section $\ref{TS_stab}$ for $\tau$ and $\sigma$. We again find that the components of $V_{ij}$ that mix real and imaginary components vanish for this solution, i.e.

\begin{equation}
 V_{ij} = \left( \begin{array}{cccccc}
V_{tt} & V_{ts} & V_{t u_{j-2}} & 0 & 0 & 0\\
V_{st} & V_{ss} & V_{s u_{j-2}} & 0 & 0 & 0\\
V_{u_{i-2} t} & V_{u_{i-2} s} & V_{u_{i-2} u_{j-2}} & 0 & 0 & 0\\
0 & 0 & 0 & V_{\tau\tau} & V_{\tau \sigma} & V_{\tau \nu_{j-2}}\\
0 & 0 & 0 & V_{\sigma\tau} & V_{\sigma\sigma} & V_{\sigma \nu_{j-2}}\\
0 & 0 & 0 & V_{\nu_{i-2} \tau} & V_{\nu_{i-2} \sigma} & V_{\nu_{i-2} \nu_{j-2}}\\
\end{array} \right)\,\,.
\end{equation}

In our approximation $\hat{\mathcal{V}} \gg \hat{\xi}$, the real components $s$ and $u_i$ are stationary points to 0-th order  of $V^{(U)}+V^{(S)}$ while $t$ is a stationary point of the first order term $V^{(T)}$ which was analyzed in section $\ref{T_stab}$. Solving

\begin{equation}
 V_s = V_{u_i} = 0
\end{equation}
for $s$ and $u_i$ is equivalent to solving eq.~\eqref{DSUiW} which we already did in section $\ref{repartsSUSYsol}$. We had found a solution by setting the flux constants $d_{1i}=d_{2i}=d_i$ so that the minima fulfilled condition eq.~\eqref{duequal}, i.e. the product $d_i u_i = d u$ for all $u_i$. We can write the scalar potential manifestly as a function of $d_i u_i$ if we write the exponential of the K\"ahler potential

\begin{equation}
 e^{K_{c.s.}} = \frac{1}{\prod_i 2 u_i} = \frac{\prod_i d_i}{\prod_i 2 d_i u_i} \equiv \frac{D}{\prod_i 2 d_i u_i}
\end{equation}
and all other terms in $V$ already appear manifestly as functions of $d_i u_i$. This simplifies the evaluation of the $(2 h^{2,1}+4) \times (2 h^{2,1}+4)$ matrix $V_{ij}$ at the stationary points $d_i u_i = d u$. After building the second derivative of $V$ with respect to at least one $d_i u_i$ we have to set the $d_i u_i = d u$. Thus, all entries of $V_{ij}$ will be mostly equal for differentiating with respect to different $u_i$, up to proportionality to $d_i$, $d_j$ or $d_i d_j$ at the stationary point. The same story holds for the axions which are set to $d_i \nu_i = 0$ after differentiating. For mixed $u_i$ and $\nu_i$ components we get

\begin{align}
 V_{t u_i} &= d_i \left( \frac{d^2 V}{dt\,d(d_i u_i)} \right)\bigg|_{d_i u_i = d u} \equiv d_i V_{t\,du}\,\,,\notag\\
 V_{s u_i} &= d_i V_{s\,du}\,\,,\notag\\
 V_{\tau \nu_i} &= d_i \left( \frac{d^2 V}{d\tau \, d(d_i \nu_i)} \right)\bigg|_{d_i \nu_i = 0} \equiv d_i V_{\tau\,d\nu}\,\,,\notag\\
 V_{\sigma \nu_i} &= d_i V_{\sigma\,d\nu}\,\,.
\end{align}

\noindent For the pure $\nu_i$ components of $V_{ij}$ we get

\begin{equation}
 V_{\nu_i \nu_j} = d_i d_j \left( \frac{d^2 V}{d(d_i \nu_i)\,d(d_j \nu_j)} \right)\bigg|_{d_i \nu_i = d_j \nu_j = 0} \equiv d_i d_j  V_{d\nu\, d\nu}
\end{equation}

\noindent which basically also holds for the pure $u_i$ components with the additional subtlety that the diagonal components $i=j$ differ from the off-diagonals $i\neq j$ which we choose to parameterize in the following way:

\begin{equation}
 V_{u_i u_j} = d_i d_j \left( \frac{d^2 V}{d(d_i u_i)\,d(d_j u_j)} \right)\bigg|_{d_i u_i = d_j u_j = d u} \equiv \begin{cases} d_i^2 (V_{d u\, d u} + \tilde{V}_{d u\, d u}) & i=j \\ d_i d_j V_{d u\, d u} & i \neq j \end{cases}\,\,.
 \label{Vuiuj}
\end{equation}

The different form of $V_{\nu_i \nu_j}$ and $V_{u_i u_j}$ can be anticipated from eq.~\eqref{VU}. $V^{(U)}$ is a function of the form $\sum_k g_k(\sum_l s_l^k x_l)$ with $s^k_l = 1$ for the $x_l$ representing the axionic fields $\nu_i$ and with

\begin{equation}
 s_l^k = \begin{cases} -1 & l=k\\ 1 & l\neq k\end{cases}
\end{equation}
for the $x_l$ representing the $u_i$. The slightly more complicated structure of $V^{(U)}$ with respect to the $u_i$ is the reason for the two different results in eq.~\eqref{Vuiuj}.

Thus, the calculation of $V_{ij}$ effectively reduces to the calculation of two matrices, one for the real parts and one for the imaginary parts of the moduli fields. The matrix for the real parts is of the following type:

\begin{equation}
\left( \begin{array}{ccccc}
V_{tt} & V_{ts} & d_1 V_{t\,du} & \dots & d_{h^{2,1}} V_{t\,du}\\
V_{st} & V_{ss} & d_1 V_{s\,du} & \dots & d_{h^{2,1}} V_{s\,du}\\
d_1 V_{t\,du} & d_1 V_{s\,du} & d_1^2 (V_{d u\, d u} + \tilde{V}_{d u\, d u}) & \dots & d_1 d_{h^{2,1}} V_{d u\, d u}\\
\vdots & \vdots & \vdots & \ddots & \vdots\\
d_{h^{2,1}} V_{t\,du} & d_{h^{2,1}} V_{s\,du} & d_{h^{2,1}} d_1 V_{d u\, d u} & \dots & d_{h^{2,1}}^2 (V_{d u\, d u} + \tilde{V}_{d u\, d u})\\
\end{array} \right)\,\,.
\label{tsu_massmatrix}
\end{equation}

\noindent For the imaginary parts, we have the same structure except the $\tilde{V}_{d \nu\, d \nu}$ term equals zero.

We now want to obtain the eigenvalues of these two matrices to 0-th order. All terms involving a $t$ or $\tau$ derivative are proportional to either $e^{-a t}$ or $\hat\xi$ and are therefore neglected.
The expressions for the eigenvalues are in general rather cumbersome so we simplify again by setting $d_i \equiv d$ for all $i$. According to section $\ref{repartsSUSYsol}$, this corresponds to demanding the VEVs of the $u_i$ to have the same values. The eigenvalues are then given by
\begin{align}
m_{1,2}^2 =& \frac{1}{2} \left[V_{ss} + (h^{2,1}  V_{uu} + \tilde{V}_{uu}) \pm \left( (V_{ss} - (h^{2,1} V_{uu} + \tilde{V}_{uu}))^2 + 4 h^{2,1} V_{su} \right)^{1/2}\right]\,\,,\label{m12realTSU}\\
m_i^2 =& \tilde{V}_{uu} = \frac{d^2 (s-1)^2}{(2u)^{h^{2,1}}\, 2s\, \gamma^2 t^3}, \qquad  \quad \text{for } i = 3,\dots,h^{2,1}+1\,\,.\label{mirealTSU}
\end{align}
Note, that we consider the masses of the moduli before canonical normalization. However, the squared masses of the canonically normalized fields will have the same overall sign as those of the unnormalized fields due to positive definiteness of the K\"ahler metric $K^{IJ}$.

\begin{figure}[h!]
\centering
\includegraphics[width=\linewidth]{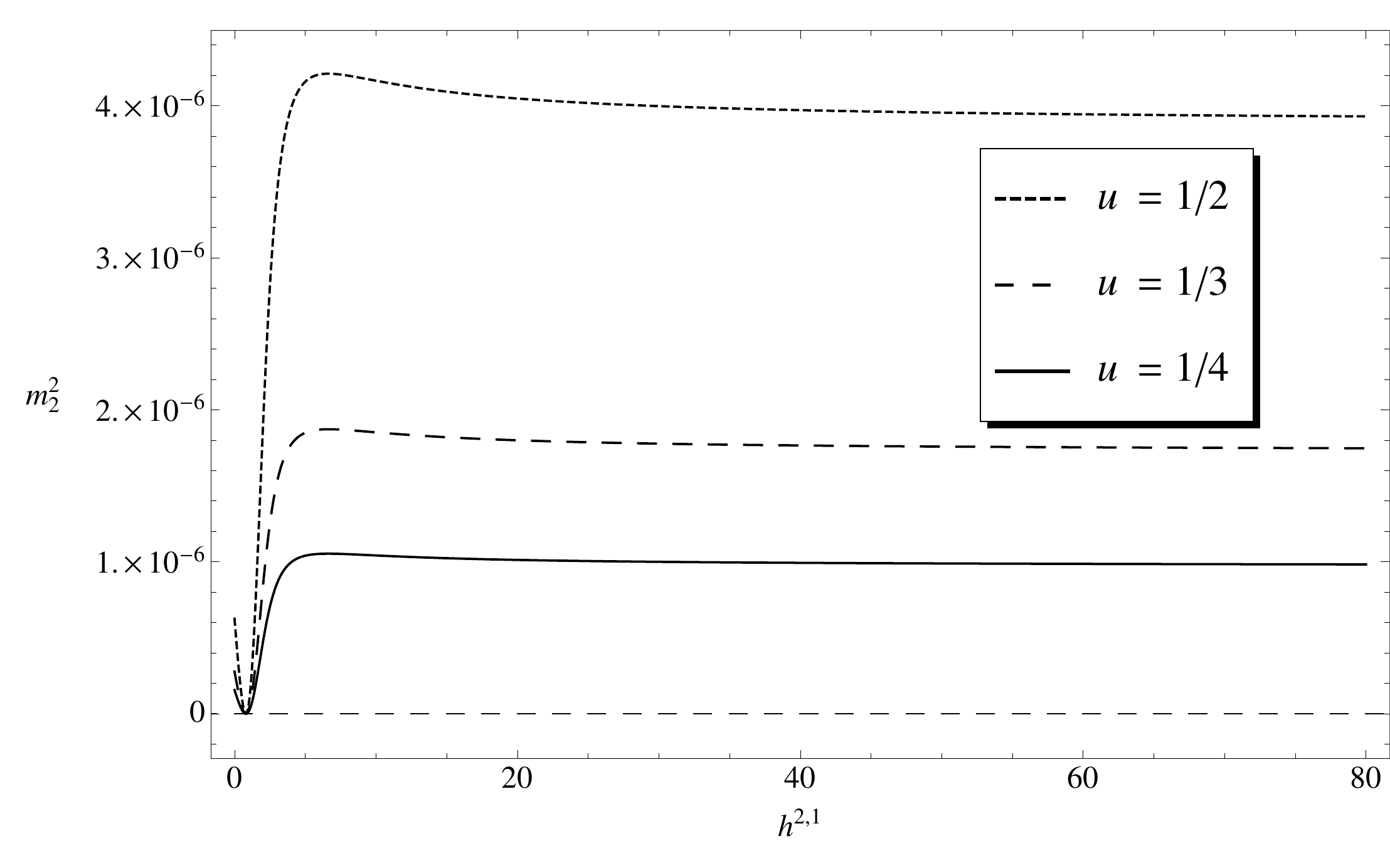}
\caption{$(2\,u)^{h^{2,1}} \cdot m_2^2$ as a function of $h^{2,1}$ for $s=10$ and three different VEVs of the complex structure moduli.}
\label{fig_m12ofh21}
\end{figure}

It is obvious that the mass $m_i^2$ in eq.~\eqref{mirealTSU} is manifestly positive whereas it is more difficult to see this analytically for $m_{1,2}^2$ since it is a rather complicated function of $s$, $u$ and $h^{2,1}$. For $m_1^2$ we find that it is a sum of two positive terms but for $m_2^2$ positivity is not obvious. However, for typical VEVs $s$ and $u$ we can plot $m_{2}^2$ as a function of $h^{2,1}$ and show that it is indeed positive, see figure $\ref{fig_m12ofh21}$. This is of course not a strict argument that $m_{1,2}^2$ is never negative. For our analysis though, it is sufficient to show that for every $h^{2,1}$ we can choose moduli VEVs which are consistent with our framework, as for instance weak string coupling, which yield positive $m_{1,2}^2$.

\noindent In the axionic sector we obtain the eigenvalues
\begin{align}
m_1^2 =&  V_{\sigma\sigma} = \frac{ d^2 u^2 [h^{2,1}+1 + 2 (h^{2,1} (h^{2,1}-1) - 1) s + (h^{2,1} (h^{2,1}-1)^2 + 1) s^2] }{(2 u)^{h^{2,1}} \,8 s^3\, \gamma^2 t^3} \,\,,\label{m1imTSU}\\[2mm]
m_2^2 =&  h^{2,1} V_{\nu\nu} = \frac{ d^2 h^{2,1} [h^{2,1} (s-1)^2+(s+1)^2] }{(2 u)^{h^{2,1}} \,8 s\, \gamma^2 t^3}\,\,, \label{m2imTSU}\\[2mm]
m_i^2 =& 0, \qquad  \quad \text{for } i = 3,\dots,h^{2,1}+1\,\,.
\end{align}

The axion masses simplify significantly due to $\tilde{V}_{\nu\nu} = V_{\sigma\nu} = 0$ at the supersymmetric minimum. Indeed, we find $h^{2,1}-1$ massless axions as we had anticipated at the end of section $\ref{impartsSUSYsol}$. The positivity of $m_1^2$ and $m_2^2$ is obvious from eq.~\eqref{m1imTSU} and eq.~\eqref{m2imTSU}.

In \cite{Conlon:2006tq} it was shown that for every unfixed axionic direction one gets a tachyonic direction in the real components of the moduli fields if all moduli are stabilized supersymmetrically. Note that this is not in contradiction with the fact that the masses eq.~\eqref{m12realTSU} and eq.~\eqref{mirealTSU} are positive which we have seen above. The crucial difference to the setting in \cite{Conlon:2006tq} is that we are not stabilizing all fields supersymmetrically, i.e. supersymmetry is spontaneously broken in the $T$ direction and so the argument of \cite{Conlon:2006tq} does not apply in our case.

The presence of massless axions per se does not constitute a serious failure of the moduli stabilization procedure, since axions couple only derivatively to all other fields at the perturbative level, and they are expected to receive a potential from non-perturbative gauge theory effects at some lower scale. A more general choice of fluxes, and/or having a CY threefold more general than $T^6$ will in general lift all of these axions, too. However, a detailed analysis of the stabilization of these complex structure moduli axions is model-dependent. Sometimes the additional non-perturbative effects which typically give them a scalar potential may not be sufficiently suppressed in scale. In such a situation the mixing of the axion masses in the total moduli mass matrix may have to be taken into account to rule out further potential accidental tachyons, which is again a model-dependent issue.

\subsection{Deviation of $s$ and $u_i$ from the SUSY minimum}

In this section, we apply the analysis of section $\ref{susybreaking_TSUgen}$ to our explicit $T^6$-based toy example for the complex structure sector eq.s~\eqref{cs_KAnsatz},\eqref{cs_WAnsatz}. Restricting ourselves again ourselves again to the case $d_i = d$, we expand around the supersymmetric 0-th order minimum $\vec{\theta}_0=(s_0,u_{0})$ for $s$ and $u_i$. Then $\delta \vec{\theta}$ is given by eq.~\eqref{deltatheta}. The matrix $V^{(0)}_{\vec{\theta}_0 \,\vec{\theta}_0}$ is written down in eq.~\eqref{tsu_massmatrix} if one eliminates the row and column that includes the derivatives with respect to $t$. Performing the matrix operations of eq.~\eqref{deltatheta} one obtains

\begin{equation}
 \delta \vec{\theta} = (\delta s, \delta u, \dots , \delta u)\,\,.
\end{equation}

\noindent With $\delta V^{(1)}_{\vec{\theta}_0} = (\delta V^{(1)}_{s_0}, \delta V^{(1)}_{u_0}, \dots, \delta V^{(1)}_{u_0})$ the components of $\delta \vec{\theta}$ are given by

\begin{align}
  \delta s &= \frac{h^{2,1} \delta V^{(1)}_{u_0} \, V_{su} - \delta V^{(1)}_{s_0}  (h^{2,1} V_{uu} + \tilde{V}_{uu})}{V_{ss} (h^{2,1} V_{u u} + \tilde{V}_{uu}) - h^{2,1}  V_{su}^2}\,\,,\label{deltas0}\\[5mm]
  \delta u &=  \frac{\delta V^{(1)}_{s_0} V_{su} - \delta V^{(1)}_{u_0} V_{ss}}{V_{ss} (h^{2,1} V_{uu} + \tilde{V}_{uu}) - h^{2,1} V_{su}^2}\,\,, \label{deltau0}
\end{align}

\noindent so essentially there is a non-trivial mixing between $\delta V^{(1)}_{s_0}$ and $\delta V^{(1)}_{u_0}$.

Now we are at the point where we can investigate the $h^{2,1}$ dependence of $\delta \vec{\theta}$. Inserting all necessary second derivatives of $V$ in eq.~\eqref{deltas0} and eq.~\eqref{deltau0} and replacing the constants $c_1$ and $c_2$ according to the supersymmetric minimum conditions eq.~\eqref{c1dus} and eq.~\eqref{c2dus} we get

\begin{align}
 \frac{\delta s}{s_0} =& -\frac{3 (s_0 - 1)^2 (47 s_0^2 - 40 s_0 + 37) }{80 (s_0^2 + 1)^2}\cdot \frac{\hat\xi}{\Vol}  + \order\left(\frac{1}{h^{2,1}}\right) \label{deltas0h21}\,\,,\\[5mm]
 \frac{\delta u}{u_0} =& -\frac{3 (s_0 - 1)^2 (47 s_0^2 - 40 s_0 + 37) }{80 (s_0^2 + 1)^2}\cdot \frac{\hat\xi}{h^{2,1} \Vol}  + \order\left(\frac{1}{(h^{2,1})^2}\right) \,\,.\label{deltau0h21}
\end{align}
%
Also terms proportional to $e^{-at}$ have been replaced using eq.~\eqref{xmintrans}. We see that $\delta s$ has a constant asymptotic behavior in $h^{2,1}$ whereas $\delta u$ decreases with $1 / h^{2,1}$. Most importantly, neither $\delta s$ nor $\delta u$ grow with positive powers of $h^{2,1}$ 
This is a very crucial point in our analysis since as mentioned above $h^{2,1} = \mathcal{O}(100)$ often appears in common realistic examples of the Calabi-Yau space like the quintic.

Of course, the specified K\"ahler potential and superpotential of eq.~\eqref{cs_KAnsatz} and eq.~\eqref{cs_WAnsatz} strictly speaking only hold for toroidal compact spaces. However, this semi-realistic construction still gives us an example for a possible $h^{2,1}$ dependence of the deviations from the supersymmetric minimum.

\subsection{Backreaction on the K\"ahler modulus} \label{kahlerback_TSUex}

We now apply the analysis of section $\ref{kahlerback_TSUgen}$ to our example eq.~\eqref{cs_KAnsatz} and eq.~\eqref{cs_WAnsatz}. We calculate $\delta V^{(2)} = V^{(T,S)} + V^{(S)} + V^{(U)}$ by expanding to first non-vanishing order in $s_0 +\delta s$ and $u_0 +\delta u$ and replacing $c_1$ and $c_2$ by the supersymmetric minimum condition eq.~\eqref{c1dus} and eq.~\eqref{c2dus}. This yields

\begin{align}
 V^{(T,S)} &\simeq -\frac{33 d (2 u_0)^{-h^{2,1}} W_0 [(s_0-1) u_0\, \delta s + h^{2,1} s_0 (s_0+1)\, \delta u]}{40 s_0^2 \Vol^2}\cdot \frac{\hat\xi}{\Vol}\,\,,\notag\\
 V^{(S)} &\simeq \frac{d^2 (2 u_0)^{-h^{2,1}} [(s_0-1) u_0\, \delta s+h^{2,1} s_0 (s_0+1)\, \delta u]^2}{2 s_0^3 \Vol^2}\,\,,\notag\\
 V^{(U)} &\simeq \frac{d^2 h^{2,1} (2 u_0)^{-h^{2,1}} [(s_0+1) u_0\, \delta s - (h^{2,1}-2) (s_0-1) s_0\, \delta u]^2}{2 s_0^3 \Vol^2} \,\,.
\end{align}

Knowing the leading order behavior of $\delta s$ and $\delta u$ in $h^{2,1}$ from eq.~\eqref{deltas0h21} and eq.~\eqref{deltau0h21}, we hence expect $V^{(T,S)}$ and $V^{(S)}$ to grow maximally $\order(1)$ and $V^{(U)}$ to grow maximally $\order(h^{2,1})$. The dependence of $\delta s$ equals the dependence of $\Delta$ on $h^{2,1}$ since $m_t^2 \propto W_0^2$ and $W_0$ is independent of $h^{2,1}$, see eq.~\eqref{W0dus}.

If we insert the values of $\delta s$ and $\delta u$ calculated in the previous section, the $\order(h^{2,1})$ contribution to $V^{(U)}$ cancels to zero. Note, that we cannot directly insert equations~\eqref{deltas0h21} and~\eqref{deltau0h21} but have to take the expressions where $e^{-a t}$ is not replaced yet. The replacement has to be performed after differentiation of $\delta V^{(2)}$ with respect to $t$ since those two operations do not commute. Hence, we are left with an $\order(1)$ expression for $\Delta$ which interestingly does not depend on $d$ and $u$, i.e. $\Delta = \Delta(s,h^{2,1})$. $\Delta$ is a rational function where the numerator and denominator are both polynomials of degree four in $s$ and degree two in $h^{2,1}$.

\begin{figure}[ht!]
\centering
\includegraphics[width=\linewidth]{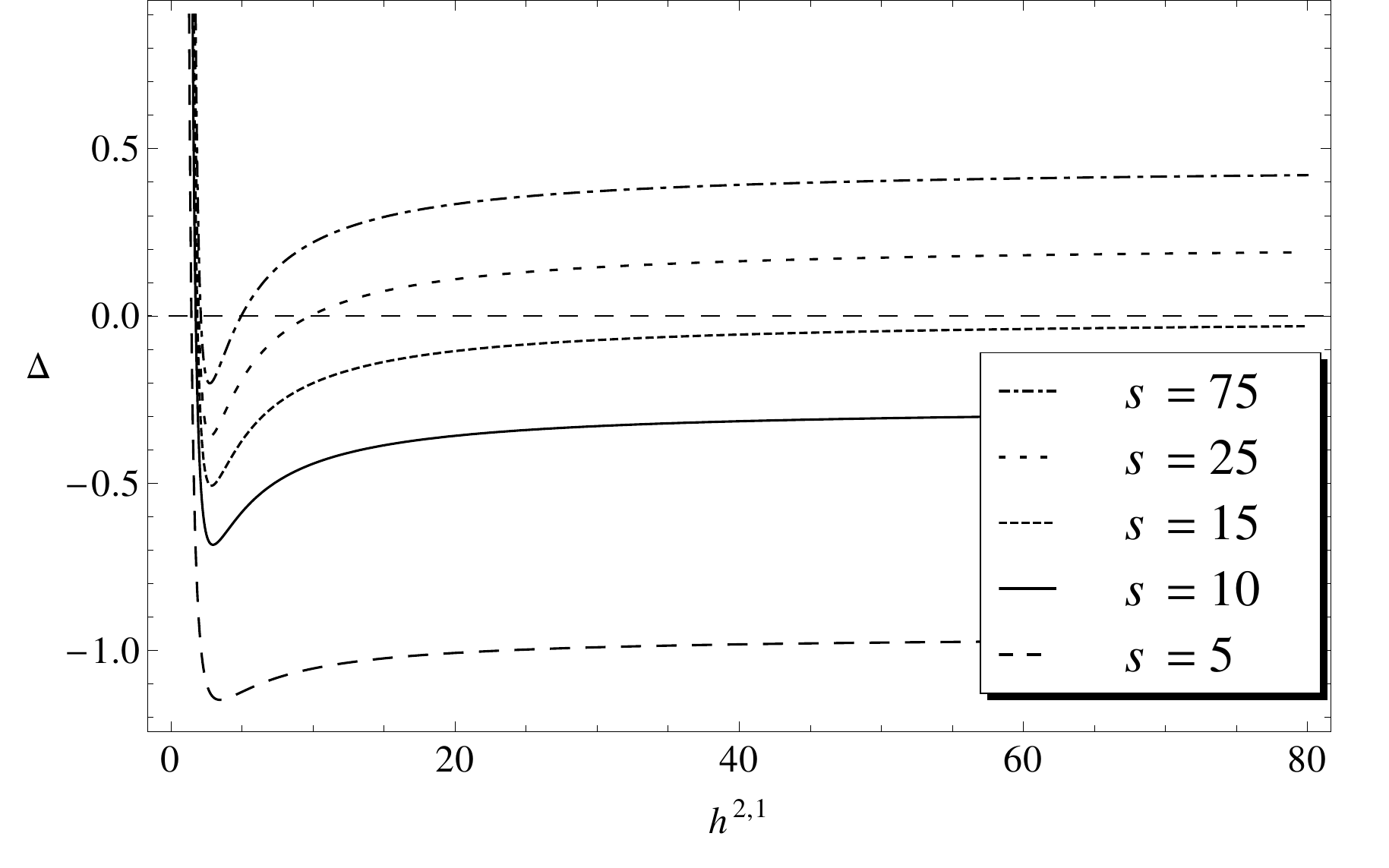}
\caption{$\Delta$ for different values of $s$.}
\label{fig_kahlerback}
\end{figure}

We plot $\Delta$ for typical values of $s$ as a function of $h^{2,1}$ in figure $\ref{fig_kahlerback}$. 
By taking the limit $s,h^{2,1} \rightarrow \infty$, one can show $\Delta < 0.6$. Furthermore, for $s \gtrsim 5$, $\Delta$ is negative with $|\Delta| < 1.2$. Hence, we conclude that $\Delta$ is always in a region where according to our analysis at the end of section $\ref{kahlerback_TSUgen}$, it does not induce a tachyonic direction in $t$.

We conclude by noting that our exemplary construction eq.~\eqref{cs_KAnsatz} and eq.~\eqref{cs_WAnsatz} has passed both potential caveats: it maintains small first-order shifts of the supersymmetrically stabilized moduli and limits the backreaction on the K\"ahler modulus VEV.

\section{Conclusions}\label{Concl}
`K\"ahler uplifting' has the benefit of generating meta-stable dS vacua in terms of just background 3-form fluxes, D7-branes and the leading perturbative ${\cal O}(\alpha'^3)$-correction, which data are completely encoded in terms of the underlying F-theory compactification on a fluxed Calabi-Yau fourfold. In addition, supersymmetry is spontaneously broken (typically $\sim M_{\rm GUT}$ here, which is below the KK scale) by an F-term generated in the volume moduli sector. No extra anti-branes, D-terms, or F-term generating matter fields are needed or involved. We emphasize that the perturbative ${\cal O}(\alpha'^3)$-correction breaks supersymmetry spontaneously. In a similar fashion as in spontaneously broken gauge theories where the ground state does not respect the full symmetry but the underlying action is invariant and thus protected from dangerous corrections by the full gauge symmetry, we expect the spontaneous breaking of supersymmetry to enhance control over the back-reaction of the various branes and orientifolds. This is because the underlying action itself still does respect the full unbroken supersymmetry. Furthermore, the backreaction of the supersymmetric D7-branes with their gauge groups is fully accounted for in the language of F-theory describing them as ADE-type singularities of the elliptic fibration of the F-theory 4-fold.

Here, we have developed a method towards a rigorous analytical understanding of `K\"ahler uplifting' driven by the leading ${\cal O}(\alpha'^3)$ correction to the K\"ahler potential of the volume moduli. Our derivation was carried out in the presence of an arbitrary number $h^{2,1}$ of complex structure moduli. A large value of 3-cycles $h^{2,1}={\cal O}(100)$  is a prerequisite to use the associated 3-form fluxes for the required fine-tuning of the cosmological constant. We have given an argument for the existence of meta-stable dS vacua in so-called `swiss cheese' type Calabi-Yau compactifications of negative Euler characteristic with an arbitrary number $h^{1,1}$ of K\"ahler moduli. The interplay of perturbative and non-perturbative effects in generating this dS minimum implies for one-parameter models with $h^{1,1}=1$ that here a structure of \emph{two} terms with alternating signs is sufficient to approximate the volume modulus scalar potential and its tunable dS vacuum. This contrasts with the `3-term structure' generically necessary in purely perturbatively stabilized situations~\cite{Saltman:2004jh,Silverstein:2007ac}. For $h^{1,1}>1$ a `3-term structure' reappears for the additional $h^{1,1}-1$ blow-up K\"ahler moduli of a `swiss cheese' Calabi-Yau.

Exploiting this `2-term structure' (or, alternatively, the `3-term structure' for $h^{1,1}>1$), we have shown that we can express the existence of the meta-stable dS vacuum for the volume modulus in terms of a \emph{sufficient} condition on the microscopic parameters. These are consisting of the fluxes, the D7-brane configuration, $h^{1,1}$, and the Euler number of the Calabi-Yau governing the perturbative ${\cal O}(\alpha'^3)$-correction, which are all in turn determined by the underlying F-theory compactification on an elliptically fibred Calabi-Yau fourfold. Thus, the result amounts to a sufficient condition for the existence of meta-stable dS vacua in terms of purely F-theory geometric and topological data which can be satisfied for a sizable subclass of all 4d ${\cal N}=1$ F-theory compactifications, instead of just single `lamp post' models.

Our sufficient condition survives both explicit inclusion of dilaton stabilization by fluxes as well as an arbitrary number of complex structure moduli.
Supersymmetry breaking happens predominantly in the volume modulus direction, and explicitly determine the shift of the dilaton and all complex structure moduli away from their flux-stabilized supersymmetric locus as suppressed by inverse powers of the volume of the Calabi-Yau. We have also checked the longevity of the metastable vacuum under tunneling.

However, there are still some possible caveats. Possible mixing with KK modes may not decouple fast enough with increasing volume of the Calabi-Yau in a given example to avoid further tachyonic directions. The structure of the 1-loop determinants of the condensing gauge groups used for volume moduli stabilization is not known in general, and it may display a dependence on the complex structure moduli which might be sufficiently strong to possibly derail our perturbative treatment of complex structure stabilization in some examples.

Finally, we have estimated the backreaction of the shifted dilaton and complex structure onto the volume modulus. The ensuing shift of the stabilized volume is generically found to be small and suppressed by inverse powers of the volume.

In conclusion, we have given arguments towards a \emph{sufficient} condition for the existence of meta-stable dS vacua in terms of, ultimately, purely F-theory data which can be satisfied for the sizable class of fluxed `swiss cheese' type Calabi-Yau compactifications with arbitrary $h^{1,1}<h^{2,1}$.

\acknowledgments We are deeply indebted and grateful to J.~Louis for many enlightening comments and discussions at many stages. We thank S. de Alwis, W.~Buchm\"uller, J.~Conlon,  L.~Covi, K. Givens, L.~McAllister, and M.~Serone for helpful comments and discussions. This work was supported by the Impuls und Vernetzungsfond of the Helmholtz Association of German Research
Centres under grant HZ-NG-603, and German Science Foundation (DFG) within the 
Collaborative Research Center 676 "Particles, Strings and the Early Universe".

\newpage
\bibliographystyle{JHEP.bst}
\bibliography{deSitter_IIB_alphaprime}
\end{document}